%% file: main.tex
\documentclass[%
 reprint,
 amsmath,
 aps, pra,
]{revtex4-2}

\usepackage{layout}

\setcitestyle{super}

\newcommand{\mytitle}{Accelerating auxiliary-field quantum Monte Carlo with mixed-precision random propagators}
\begin{document}

\preprint{}

\title{\mytitle}

\author{Jacob Z.~Williams\,\orcidlink{0000-0002-8467-9211}}
\affiliation{%
  Theoretical Division, Los Alamos National Laboratory,\\
  Los Alamos, New Mexico 87545, United States
}%

\author{Yu Zhang\,\orcidlink{0000-0001-8938-1927}}
  \email{zhy@lanl.gov}
\affiliation{%
  Theoretical Division, Los Alamos National Laboratory,\\
  Los Alamos, New Mexico 87545, United States
}%

\date{\today}

\begin{abstract}
\input{kernel_abstract}
\end{abstract}

\maketitle

\input{kernel_main}

\begin{acknowledgments}
\input{kernel_acknow}
\end{acknowledgments}

\bibliography{mp-afqmc}

\clearpage

\widetext
\begin{center}
\textbf{\large%
Supporting Information for ``\mytitle''
}
\end{center}
\setcounter{section}{0}
\setcounter{equation}{0}
\setcounter{figure}{0}
\setcounter{table}{0}
\setcounter{page}{1}
\makeatletter
\renewcommand{\thesection}{S\arabic{section}}
\renewcommand{\theequation}{S\arabic{equation}}
\renewcommand{\thefigure}{S\arabic{figure}}
\renewcommand{\bibnumfmt}[1]{[S#1]}
\renewcommand{\citenumfont}[1]{S#1}

\input{kernel_supp}

\end{document}

%% file: kernel_abstract.tex
We develop an algorithm to propagate auxiliary-field quantum Monte Carlo (AFQMC)
wavefunctions in mixed precision.
We target NVIDIA Tensor Cores using the TensorFloat-32 (TF32) format.
We approximate each full-precision array
by a TF32 leading term and a TF32 residual,
and each full-precision matrix product by three TF32 products
accumulated in full precision.
Benchmarks on hydrogen chains show that mixed-precision propagation
is consistently faster than double-precision GPU propagation,
with an increasing advantage as the system size grows.
The resulting local-energy deviations remain small
over the propagation windows considered here.
These results identify the stochastic two-body propagator
as a particularly effective target for reduced precision,
while our method allows higher precision to be retained
for accuracy-sensitive operations.
Hence, our algorithm provides a practical way to speed up AFQMC calculations
and demonstrates where reduced precision can be applied
without adverse effects on the method's accuracy.

%% file: kernel_main.tex
\section{\label{sec:intro}Introduction}
The electronic structure problem---accurately computing the energy and other
ground-state properties of correlated electronic systems---is a critical part
of modern chemistry and materials science research
\cite{greeley_etal_2002, marzari_etal_2021}.
The scale of its impact is difficult to overstate; as of 2025,
three of the ten most cited papers of all time~\cite{vannoorden_2025}
describe electronic structure theory methods 
\cite{lee_etal_1988, perdew_etal_1996a, kresse_furthmuller_1996}.
All three apply density functional theory (DFT),
known as the ``workhorse of quantum chemistry"~\cite{teale_etal_2022},
but methods beyond DFT are required to describe strongly correlated systems
accurately enough for cases like
(nearly) degenerate spin configurations in transition metals~\cite{neese_2006}
and Mott insulators~\cite{pavarini_2021}.
Their usefulness is due in large part to the robust ecosystem of
electronic structure software, especially when coupled with
high-performance computing (HPC) hardware~\cite{gavini_etal_2023}.
Looking forward, it is crucial that electronic structure codes and algorithms
are tailored to the fastest HPC hardware available.
This hardware is increasingly shaped by the unique requirements
of machine learning (ML) methods.
On one hand, an increasing fraction of HPC time is spent on ML computation
\cite{bianchini_etal_2026}; on the other, many cutting-edge HPC systems
are now purpose-built for ML~\cite{resch_etal_2025}.
ML models are often run in reduced-precision arithmetic~\cite{gupta_etal_2015}
on graphics processing units (GPUs)
or specialized hardware accelerators
with exceptional performance in low precision~\cite{dean_2020}.
It is therefore important to update electronic structure algorithms to use
mixed-precision arithmetic to take full advantage
of modern HPC and ML hardware~\cite{shinde_etal_2025a}.

Auxiliary-field quantum Monte Carlo (AFQMC) is a promising method for accurate
electronic structure calculations on correlated systems
\cite{shi_zhang_2021, lee_etal_2022},
and is ideally suited for the heterogeneous hardware architectures
likely to dominate the HPC ecosystem in the near- and medium-term future.
AFQMC transforms the two-body interaction term of the Hamiltonian
into an ensemble of one-body potentials weighted by probabilistic auxiliary fields.
The auxiliary fields are sampled stochastically, and each realization is used
to propagate an independent random walker through imaginary time.
AFQMC can recover energetics with accuracy approaching that of
``gold-standard" coupled-cluster theory,
with improved asymptotic efficiency~\cite{lee_etal_2022, shee_etal_2023}.
However, AFQMC still suffers significant bottlenecks in practice;
for systems of moderate size, propagating random walkers through imaginary time
dominates the computational cost.
We address this bottleneck by applying the auxiliary-field propagators
on GPUs in mixed precision.

AFQMC is embarrassingly parallel~\cite{church_rubenstein_2021, shee_etal_2023}
and naturally suited for execution on GPUs~\cite{shee_etal_2018};
each random walker can be allocated to a separate processor if necessary.
In addition, replacing the two-body interaction
by fluctuating auxiliary fields induces statistical uncertainty,
so reduced precision may have a smaller effect on propagation
under the auxiliary fields than on deterministic operations.
Indeed, the electron--electron interaction is typically
approximated by a Cholesky decomposition
\cite{roeggen_wisloff-nilssen_1986, pedersen_etal_2024}
even before it is recast into a sum of fluctuating fields,
mitigating the impact of reduced precision for the two-body propagation further.
We note that~\citet{shee_etal_2018} have already investigated the use of
reduced (single) precision propagation,
obtaining a substantial speedup, with errors of roughly \qty{1}{\milli\hartree}.
Here, we approximate full-precision operations by several low-precision operations
to exploit the higher throughput of Tensor Cores
while recovering much of the precision lost in direct TF32 propagation.

We introduce the AFQMC algorithm and present our mixed-precision
propagation scheme in Section~\ref{sec:theory};
demonstrate its accuracy--speed tradeoff in Section~\ref{sec:results};
and discuss future directions in Section~\ref{sec:summary}.

\section{\label{sec:theory}Methodology}

\subsection{Auxiliary-field quantum Monte Carlo} \label{sec:theory-afqmc}
We here describe (phaseless) auxiliary-field quantum Monte Carlo, or (ph-)AFQMC,
briefly. For a more detailed discussion,
we refer the reader to previous reviews~\cite{motta_zhang_2018, shi_zhang_2021}.
AFQMC projects an initial trial state $\ket{\Psi_{\trial}}$
through imaginary time $\tau$, obtaining the true ground state
\begin{equation}
    \ket{\Psi_0} = \lim_{\tau \to \infty} e^{-\tau \Ham} \ket{\Psi_{\trial}}
\end{equation}
as long as the trial and ground states are not orthogonal.
It exploits the Hubbard--Stratonovich transformation
\cite{stratonovich_1957, hubbard_1959}
to transform the electron--electron interaction
into many fluctuating one-body potentials,
thus changing the many-body problem
into a sum of probabilistic one-body auxiliary fields,
which admits a mean-field imaginary time evolution.
The total AFQMC wavefunction is a weighted sum of walker wavefunctions $\ket{\phi_w}$
that evolve under independent realizations of the auxiliary fields,
\begin{equation}
    \ket{\Psi_\mc} = \sum_w c_w \ket{\phi_w}.
\end{equation}
In this work, we take $\ket{\Psi_\trial}$,
which doubles as the initial state of each walker,
as a single Slater determinant.
In this setting, Thouless' theorem~\cite{thouless_1960}
guarantees that the walkers remain single Slater determinants throughout propagation,
allowing use of efficient numerical routines very similar to those
for Hartree--Fock or Kohn--Sham density functional theory.
The many-body correlations are recovered in expectation
by the ensemble average in $\ket{\Psi_\mc}$.

\subsubsection{The AFQMC Hamiltonian}
The electronic structure Hamiltonian in second quantization is
\begin{equation}
    \Ham = \Op{h} + \Op{V}_{\elec \elec} + C,
\end{equation}
where $\Op{h} = \sum_{ij} h_{ij} \cdag_i \cdes_j$ is the one-body term,
$\Op{V}_{\elec \elec} = \frac12 \sum_{ijkl} v_{ijkl} \cdag_i \cdag_j \cdes_k \cdes_l$
is the electron--electron interaction, and $C$ is a constant energy shift. We
take the Born--Oppenheimer approximation, so $C$ includes for example the
electrostatic repulsion between nuclei.

In order to apply the Hubbard--Stratonovich transformation and use the machinery
of AFQMC, we recast $\Op{V}_{\elec\elec}$ into a sum of squared one-body terms,
\begin{equation}
    \Op{V}_{\elec\elec} = \frac12 \sum_\gamma \Op{L}_\gamma^2,
\end{equation}
where $\Op{L}_\gamma = \sum_{ij} L_{\gamma,ij} \cdag_i \cdes_j$.
Since the Coulomb operator is positive semidefinite, this can be accomplished by
the (modified) Cholesky decomposition
\cite{cholesky_1924, beebe_linderberg_1977, koch_etal_2003, 
roeggen_johansen_2008, aquilante_etal_2010}:
\begin{equation}
\begin{split}
    v_{ijkl} = & v_{(il),(jk)}
    = \sum_\gamma U_{kj,\gamma} D_\gamma U_{\gamma, il}^\dagger\\
    = &\sum_\gamma \left(\sqrt{D_\gamma} U_{kj, \gamma} \right)
        \left( \sqrt{D_\gamma} U_{il, \gamma}^* \right)\\
    = &\sum_\gamma L_{kj, \gamma} L_{il, \gamma}^*.
\end{split}
\end{equation}
A low-rank approximation is then introduced,
neglecting the eigenvalues $D_\gamma$ less than a cutoff $\alpha_L$.
Unless otherwise indicated, we set $\alpha_L = 10^{-6}$.
The number of Cholesky matrices required to represent $\Op{V}_{\elec\elec}$
to a fixed accuracy $\alpha_L$ is linear in the number $N_\AO$ of basis functions
\cite{roeggen_wisloff-nilssen_1986, pedersen_etal_2024},
reducing the total storage required for the electron repulsion integrals
from $\Ord{(N_\AO^4)}$ to $\Ord{(N_\AO^3)}$.
It can be further reduced to $\Ord{(N_\AO^2)}$
by leveraging tensor hypercontraction~\cite{hohenstein_etal_2012,parrish_etal_2012,hohenstein_etal_2012a}
or mixed sparsity--double factorization methods~\cite{Zhang:2026aa}.
Using the Cholesky decomposition, the Hamiltonian can be expressed
in the so-called MC form,
with a one-body operator and a sum of squared one-body operators,
\begin{equation} \label{eq:ham-mc}
    \Op{H}_{\mc} = \Op{T} + \frac12 \sum_\gamma \Op{L}_\gamma^2 + C,
\end{equation}
where
\begin{equation}
    \Op{T} = \sum_{ij} 
        \left[
            h_{ij} - \frac12 \sum_\gamma \sum_k L_{ik,\gamma} L_{jk,\gamma}^* 
        \right]
        \cdag_i \cdes_j,
\end{equation}
is the shifted kinetic operator.

\subsubsection{Discretization of imaginary time propagation}
In order to implement $\lim_{\tau \to \infty} e^{-\tau \Ham}$ numerically,
we fix a maximum imaginary time $\tau_{\mathrm{max}}$ and discretize the propagation
into $N_\tau$ steps of size $d\tau$ ($N_\tau = \tau_{\mathrm{max}}/d\tau$):
\begin{equation}
    \lim_{\tau \to \infty} e^{-\tau \Ham_{\mc}} \approx
    \left(e^{-d\tau\,\Ham_{\mc}} \right)^{\tau_{\mathrm{max}}/d\tau}.
\end{equation}
Performing a symmetric second-order Lie--Suzuki--Trotter
decomposition~\cite{trotter_1959, suzuki_1976} on $\Ham_{\mc}$, we have
\begin{multline}
    e^{-d\tau\, \Ham_{\mc}} = \\
    e^{-d\tau/2\, \Op{T}} e^{-d\tau\, \sum_\gamma \Op{L}_\gamma^2/2} 
        e^{-d\tau/2\, \Op{T}} e^{-d\tau\, C} + \Ord{(d\tau^3)},
\end{multline}
incurring error cubic in $d\tau$.

\subsubsection{Propagation and Monte Carlo sampling}
Once $\Op{V}_{\elec\elec}$ is cast into the Monte Carlo form in Eq.~\eqref{eq:ham-mc},
its corresponding propagator can be decomposed into an integral
over one-body propagators by the Hubbard--Stratonovich (HS) transformation
\cite{stratonovich_1957, hubbard_1959} as
\begin{equation}
    e^{-d\tau\,\Op{L}_\gamma^2/2} =
    \int dx_\gamma\, \frac{e^{-x_\gamma^2/2}}{\sqrt{2\pi}} e^{x_\gamma\sqrt{-d\tau} \Op{L}_\gamma},
\end{equation}
where each $x_\gamma$ is an auxiliary field, here taking the form of a
standard normal random variable (so that the probability of $x_\gamma$ taking the
value $x$ is $p(x) = e^{-x^2/2}/\sqrt{2\pi}$).
Performing this transformation on $\Op{V}_{\elec\elec}$,
we obtain one auxiliary field per Cholesky matrix;
the operator that propagates the wavefunction by $d\tau$ is
\begin{widetext}
\begin{equation}
    e^{-d\tau\, \Ham_{\mc}} =
    \int \cdots \int 
    \prod_\gamma dx_\gamma\, \frac{e^{-x_\gamma^2/2}}{\sqrt{2\pi}} \times
    e^{-d\tau/2\, \Op{T}}
        \exp{\left(\sqrt{-d\tau}\, \sum_\gamma x_\gamma \Op{L}_\gamma\right)} 
        e^{-d\tau/2\, \Op{T}}
        e^{-d\tau\, C}
    + \Ord{(d\tau^2)}.
\end{equation}
\end{widetext}
By placing the sum $\sum_\gamma x_\gamma \Op{L}_\gamma$ in one exponential,
we incur an increase in the Trotter error from cubic to quadratic in $d\tau$.

Now expressed as a high-dimensional stochastic integral,
imaginary-time propagation is amenable to Monte Carlo sampling.
We sample one independent realization of each $x_\gamma$ at each time step
on an ensemble of walker wavefunctions $\ket{\phi_w}$.
Each walker starts in the trial state $\ket{\Psi_\trial}$,
but evolves differently from the others due to the randomness in $\{x_\gamma\}$.
Averaging over the walker ensemble recovers the many-body correlation: that is,
we replace many-body dynamics by a fluctuating ensemble of one-body propagations.
Observable estimates have a standard deviation proportional to the
inverse square root of the number of random walkers~\cite{motta_zhang_2018}.

In order to improve the efficiency of sampling $x_\gamma$
with respect to a given trial wavefunction $\ket{\Psi_\trial}$,
a kind of importance sampling is introduced~\cite{zhang_etal_1997}.
Importance sampling biases the auxiliary fields depending on the
expected value of $\Op{L}_\gamma$ with respect to $\ket{\Psi_\trial}$,
\begin{equation}
    \frac{e^{-x_\gamma^2/2}}{\sqrt{2\pi}} \mapsto 
    \frac{e^{-(x_\gamma - F)^2/2}}{\sqrt{2\pi}} \times
        e^{\frac{F^2}{2} - x_\gamma F}
\end{equation}
where the so-called dynamic force
\begin{equation}
    F = \sqrt{-d\tau}\, 
    \frac{\mel{\Psi_\trial}{\Op{L}_\gamma}{\phi_w}}
         {\braket{\Psi_\trial}{\phi_w}}.
\end{equation}
The latter term, $\exp{\left(F^2/2 - x_\gamma F \right)}$,
is a normalization factor.
We apply this bias in our mixed-precision propagation,
but to save space we write $x_\gamma$ for the biased realizations.

\subsubsection{The phaseless constraint}
In practice, AFQMC incurs a phase problem,
akin to but even more severe than the sign problem~\cite{loh_etal_1990, ceperley_1991}
of other QMC methods.
The complex propagators cause the walker wavefunctions
to accumulate fluctuating phases,
with variances increasing exponentially with the number of steps $d\tau$.
To remedy this, we apply the phaseless constraint of~\citet{zhang_krakauer_2003},
which uses two strategies to mitigate the phase problem.
First, the importance-sampling transformation described above shifts the auxiliary fields using the trial wavefunction.
Second, the complex weights $c_w$ of each walker wavefunction $\ket{\phi_w}$
are projected to the positive real axis at each step.
Together, the phaseless constraint controls the phase problem to
$\Ord{(\sqrt{d\tau})}$,
but introduces a fundamental constraint on the accuracy.
Unless $\ket{\Psi_\trial}$ is the many-body ground state,
phaseless AFQMC is no longer exact, even in the limit $\tau \to \infty$.
Furthermore, because of the bias induced by the trial wavefunction,
the energy of ph-AFQMC is no longer variational~\cite{zhang_krakauer_2003}.
Extensive research efforts have focused on developing improved trial wavefunctions
for AFQMC~\cite{shee_etal_2018, shi_zhang_2021, danilov_etal_2025, jiang_etal_2025a, mahajan_etal_2025, xiao_etal_2025, xiao_etal_2025a, yoshida_etal_2025, williams2026df};
in this work, we take $\ket{\Psi_\trial}$
to be a restricted Hartree--Fock wavefunction.

\subsubsection{Measuring the energy}
One-body operators $\Op{O}$ that commute with the Hamiltonian
(including, of course, the one-particle energy operator $\Op{T}$)
are measured by the mixed estimator
\begin{equation} \label{eq:mixed-onebody}
    \evo{\Op{O}} 
        = \frac{\sum_w c_w \mel{\Psi_\trial}{\Op{O}}{\phi_w}}
               {\sum_w c_w \braket{\Psi_\trial}{\phi_w}},
\end{equation}
where $\mel{\Psi_\trial}{\Op{O}}{\phi_w}$ is known as a (per-walker)
local observable \cite{shi_zhang_2021}.
This estimator is exact when $\Op{O}$ commutes with the Hamiltonian;
otherwise, it generally retains a bias from the trial state.
Two-body operators like $\Op{V}_{\elec\elec}$ require the use
of the generalized Wick theorem~\cite{wick_1950, balian_brezin_1969}:
since $\ket{\Psi_\trial}$ and each walker wavefunction $\ket{\phi_w}$
are Slater determinants, the per-walker equal-time two-body Green's function
\begin{equation}
    G^{(w)}_{ij\sigma,kl\tau} =
        \frac{
            \mel{\Psi_\trial}
                {\cdag_{i\sigma} \cdag_{j\tau} \cdes_{k\tau} \cdes_{l\sigma}}
                {\phi_w}
             } 
             {\braket{\Psi_\trial}{\phi_w}}
\end{equation}
decomposes in terms of the one-body Green's functions
\begin{equation}
    G^{(w)}_{ij\sigma} = 
        \frac{\mel{\Psi_\trial}{\cdag_{i\sigma} \cdes_{j\sigma}}{\phi_w}}
             {\braket{\Psi_\trial}{\phi_w}}
\end{equation}
as
\begin{equation}
    G^{(w)}_{ij\sigma,kl\tau} = 
        G^{(w)}_{il\sigma} G^{(w)}_{jk\tau} - 
        \delta_{\sigma\tau} G^{(w)}_{ik\sigma} G^{(w)}_{jl\sigma}.
\end{equation}
$G^{(w)}_{ij\sigma}$ can be computed efficiently in terms of the Slater
determinant matrices of $\ket{\Psi_\trial}$ and $\ket{\phi_w}$.

Thus the local two-body energy of walker $w$ is
\begin{widetext}
\begin{equation} \label{eq:locale2}
\begin{split}
    \mel{\Psi_\trial}{\Op{V}_{\elec\elec}}{\phi_w} &= 
        \frac12 \sum_\gamma \sum_{ijkl} \sum_{\sigma\tau}
            L_{il, \gamma} \conj{L}_{jk, \gamma}
        \frac{ 
                \mel{\Psi_\trial}
                    {\cdag_{i\sigma} \cdag_{j\tau} \cdes_{k\tau} \cdes_{l\sigma}}
                    {\phi_w}
             }
             {\braket{\Psi_\trial}{\phi_w}} \\ &=
        \frac12 \sum_\gamma \sum_{ijkl} \sum_{\sigma\tau}
            L_{il,\gamma} \conj{L}_{jk, \gamma}
        \left( G^{(w)}_{il\sigma} G^{(w)}_{jk\tau} - 
        \delta_{\sigma\tau} G^{(w)}_{ik\sigma} G^{(w)}_{jl\sigma} \right),
\end{split}
\end{equation}
\end{widetext}
and the mixed estimator is
\begin{equation}
    \evo{\Op{V}_{\elec\elec}} = 
        \frac{\sum_w c_w \mel{\Psi_\trial}{\Op{V}_{\elec\elec}}{\phi_w}}
             {\sum_w c_w \braket{\Psi_\trial}{\phi_w}}.
\end{equation}

The first term on the right-hand side of \eqref{eq:locale2}
corresponds to the Hartree energy and is easy to compute.
The second term corresponds to exchange.
Requiring $\Ord{(N_\AO^4)}$ operations asymptotically,
\cite{motta_zhang_2018, motta_etal_2019a, Zhang:2026aa}
it becomes the bottleneck of AFQMC.
However, the energy does not need to be computed at every propagation step.
For systems of moderate size, the propagation
(and in particular the memory overhead for storing the Cholesky matrices $\{\Op{L}_\gamma\}$)
are the practical bottlenecks.

\subsection{Reduced-precision arithmetic} \label{sec:theor-redprec}
Typical scientific computation workflows store real numbers,
or the real and imaginary parts of complex numbers,
as 64-bit binary floating-point numbers (FP64).
According to the IEEE 754 standard~\cite{ieee_2019},
a floating-point number $f$ is expressed as
\begin{equation}
    f = 
    (-1)^{s} \times 
        2^{e - (2^{N_e - 1} - 1)} \times 
        \sum_{n=1}^{N_b} b_n 2^{-(n-1)}.
\end{equation}
It is stored on a computer in binary as $s e_1 \dots e_{N_e} b_1 \dots b_{N_b}$.
The first bit $s$ encodes the sign;
an $N_e$-bit exponent is encoded in $e = e_1 \dots e_{N_e}$,
noting that the offset in the exponent allows both positive and negative exponents
up to $2^{N_e - 1} - 1$ to be represented;
and the significand is encoded in $N_b$ bits $b_i$,
where biasing the exponent yields one additional significand bit implicitly.
FP64 is defined by setting $N_e = 11$ and $N_b = 52$.
It represents numbers up to $\pm 2^{1023} \approx 10^{300}$
with a relative precision (the 64-bit machine epsilon) of
$\eps_{\dfp} = 2^{-53} \approx 1.1 \times 10^{-16}$.

FP64 has enjoyed decades of algorithmic development
since its standardization in 1985~\cite{ieee_1985}.
It is a battle-tested data structure for numerical and scientific computing,
especially on central processing units (CPUs).
Recent years have seen a boom in graphics processing unit (GPU) computing,
especially ML applications,
that favor reduced-precision floating-point numbers. For example, IEEE 754
also defines the 32-bit FP32, which has $N_e = 8$, $N_b = 23$;
FP32 can represent numbers up to about $\pm 1.7 \times 10^{38}$, 
and $\eps_\sfp = 2^{-24} \approx 6 \times 10^{-8}$.
More recently, NVIDIA introduced the TensorFloat-32 (TF32) format
for its Ampere series GPUs~\cite{stosic_micikevicius_2021},
with $N_e = 8$, $N_b = 10$.
TF32 represents the same exponent range as FP32,
but with precision further reduced to
$\eps_\tfp = 2^{-11} \approx 4.9 \times 10^{-4}$.

A na\"ive estimate, roughly correct for the NVIDIA A100 GPU,
is that double-precision (FP64) floating point operations (FLOPs)
take twice as long as single-precision (FP32) ones.
The A100's architecture, however, is designed around the TF32 format,
executing operations in TF32 eight times faster, at $156 \times 10^{12}$ floating-point
operations per second (\qty{156}{\tera\FLOP\per\second}), than in FP32
(\qty{19.5}{\tera\FLOP\per\second}),
and sixteen times faster than in FP64 (\qty{9.7}{\tera\FLOP\per\second})
\cite{nvidia_a100}.

Mixed-precision arithmetic,
in which some operations are computed at less than FP64 precision,
has been used with success in quantum chemical calculations,
from density matrix purification for semiempirical~\cite{steinbach_bannwarth_2025}
and density functional~\cite{woo_choi_2026} calculations
to many-body methods like M{\o}ller--Plesset
and coupled-cluster theory~\cite{olivares-amaya_etal_2010,eriksen_2017}
and the density matrix renormalization group~\cite{tian_etal_2022}.
Accelerating AFQMC on graphics processing units has been investigated
in double (FP64)~\cite{huang_etal_2024}
and single (FP32) precision~\cite{shee_etal_2018},
producing substantial speedups and (for FP32) errors below the
\qty{1.6}{\milli\hartree} threshold of chemical accuracy.
Our mixed-precision algorithm uses two TF32 numbers to represent one
full-precision number. We can recover accuracy close to
the single-precision propagation of \citet{shee_etal_2018}
while improving speed even further on GPUs with Tensor Cores.

\citet{thall_2006} discusses bitwise operations for adding and
multiplying an FP64 number represented as two FP32 numbers,
with an effective precision of roughly 48 bits. 
We use a similar splitting technique, Algorithm~\ref{alg:split-float},
to approximate an FP64 number with two TF32 floats,
$f_\tfp^\hlprec = (f_\tfp^\hprec, f_\tfp^\lprec)$;
the same algorithm would work for FP32.
For TF32, with 10 significant bits,
the effective precision of Algorithm~\ref{alg:split-float}
is about 20 bits, nearly the 23 bits of FP32.
(We implicitly assume that $f_\dfp$
lies in the exponent range expressible by TF32.)

\begin{algorithm}[htb]
    \algcaption{Approximate an FP64 float as two TF32 floats.}
    \label{alg:split-float}
    \begin{algorithmic}[1]
        \Procedure{SplitPrecision}{$f_\dfp$}
        \State $f_\tfp^\hprec \gets \FPT{(f_\dfp)}$
        \Comment{Or $\sfp$, throughout}
        \State $f_\tfp^\lprec \gets \FPT{\left(f_\dfp - \FPD{(f_\tfp^\hprec)}\right)}$
        \State \textbf{return} $f_\tfp^\hprec$, $f_\tfp^\lprec$
        \EndProcedure
    \end{algorithmic}
\end{algorithm}

Neglecting under- and overflow, adding two numbers
represented in this fashion is simply adding the components,
\begin{equation}
    (f + g)_\tfp^\hlprec =
    (f_\tfp^\hprec + g_\tfp^\hprec, f_\tfp^\lprec + g_\tfp^\lprec).
\end{equation}

Following \citet{valero-lara_etal_2023}, we approximate (matrix) multiplication
by three reduced-precision multiplications (Algorithm~\ref{alg:mult}).
This extends the multiword algorithm of \citet{fasi_etal_2023} to the TF32 datatype
and is similar to the Ozaki scheme for exact emulation
of higher-precision arithmetic~\cite{ozaki_etal_2012},
while neglecting the product of the low bits
in the spirit of a first-order Taylor expansion.
As the exponent ranges of TF32 and FP32 are smaller than that of FP64,
we assume that the product $f_\tfp^\hprec g_\tfp^\hprec$
can be represented in reduced precision;
our numerical tests have not yielded over- or underflow errors.

\begin{algorithm}[htb]
    \algcaption{Approximately multiply two FP64 structures 
                using TF32 operations.}
    \label{alg:mult}
    \begin{algorithmic}[1]
        \Procedure{MixPrecMult}{$f_\dfp, g_\dfp$}
        \State $f_\tfp^\hprec, f_\tfp^\lprec \gets
            \operatorname{\textsc{SplitPrecision}}{(f_\dfp)}$
        \State $g_\tfp^\hprec, g_\tfp^\lprec \gets
            \operatorname{\textsc{SplitPrecision}}{(g_\dfp)}$
        \State \textbf{return} 
            $\FPD{\left(%
                f_\tfp^\hprec g_\tfp^\hprec + 
                f_\tfp^\hprec g_\tfp^\lprec + 
                f_\tfp^\lprec g_\tfp^\hprec
            \right)}$
        \EndProcedure
    \end{algorithmic}
\end{algorithm}

\subsection{Mixed-precision AFQMC propagation}
We approximate the Monte Carlo propagation on each walker,
\begin{equation}
    \ket{\phi_w} \to 
    e^{-\frac{d\tau}{2}\, \Op{T}} e^{\sqrt{-d\tau} 
    \sum_\gamma x_\gamma \Op{L}_\gamma} 
    e^{-\frac{d\tau}{2}\, \Op{T}} 
        \ket{\phi_w},
\end{equation}
with Algorithms \ref{alg:split-float} and \ref{alg:mult},
representing high and low bits of $\ket{\phi_w}$,
$\sqrt{-d\tau}\, \sum_\gamma x_\gamma \Op{L}_\gamma$, and optionally
$e^{-\frac{d\tau}{2}\, \Op{T}}$ in reduced precision.

We note that, while $\Op{T}$ and the trial wavefunction $\ket{\Psi_\trial}$
are purely real for molecular systems, 
the mean-field shift means that the modified two-body propagator
is complex, so $\ket{\phi_w}$ acquires an imaginary component.
Nevertheless, we can batch
$\Real{x_\gamma} L_\gamma$ and $\Imag{x_\gamma} L_\gamma$
into a single real array and multiply both by $\ket{\phi_w}$,
inserting the imaginary unit $i$ afterward.
We thus save time relative to a na\"ively complex matrix multiplication,
which requires four real matrix-matrix multiplications.
If $\Op{T}$, $\Op{L}_\gamma$, or $\ket{\phi_w}$ are intrinsically complex,
as is the case in periodic boundary conditions,
this decomposition is required:
\begin{equation}
\label{eq:matmul-cplx}
\begin{split}
    \Op{A} \ket{\phi} &= 
    (\Op{A}_r + i \Op{A}_i) (\ket{\phi}_r + i \ket{\phi}_i ) \\ &=
    (\Op{A}_r \ket{\phi}_r - \Op{A}_i \ket{\phi}_i) + 
        i (\Op{A}_r \ket{\phi}_i + \Op{A}_i \ket{\phi}_r).
\end{split}
\end{equation}
Since each term in \eqref{eq:matmul-cplx} is approximated
by Algorithm~\ref{alg:mult} by three reduced-precision multiplications,
the complex product $\Op{A}\ket{\phi}$
requires twelve real matrix products
(and two real accumulations in full precision).
We leave this fully complex implementation for future work.

We further approximate the action of the two-body propagator
$e^{\sqrt{-d\tau}\,\sum_\gamma x_\gamma \Op{L}_\gamma}$ on $\ket{\phi_w}$
by a recursive procedure implementing the Taylor expansion
of the exponential, Algorithm~\ref{alg:exptay}.
This is to avoid having to compute an expensive matrix exponential
for every walker at every timestep;
while the Cholesky matrices $\Op{L}_\gamma$ do not change,
a different realization of each $x_\gamma$ is required for each propagation step
and each independent walker.
($\frac{-d\tau}{2}\, \Op{T}$ has to be exponentiated only once,
at the start of propagation; we thus exponentiate it exactly.)

\begin{algorithm}[htb]
    \algcaption{$e^{\Op{A}} \ket{\psi}$ approximated
                by order-$N$ Taylor expansion.}
    \label{alg:exptay}
    \begin{algorithmic}[1]
        \Procedure{ExpTaylor}{$\Op{A}, \ket{\psi}, N$}
        \State $\ket{\pi} \gets \ket{\psi}$
            \Comment{Product accumulator $\Op{A}^n\ket{\psi}/n!$}
        \State $\ket{\sigma} \gets \ket{\psi}$
            \Comment{Sum accumulator; $\Op{A}^0 \ket{\psi}/0! = \ket{\psi}$}
        \For{$n$ from $1$ to $N$}
            \State $\ket{\pi} \gets \Op{A} \ket{\pi} / n$
            \State $\ket{\sigma} \gets \ket{\sigma} + \ket{\pi}$
        \EndFor
        \State \textbf{return} $\ket{\sigma}$
        \EndProcedure
    \end{algorithmic}
\end{algorithm}

We apply the two-body propagator in mixed precision,
accumulating the real and imaginary components
of the updated $\ket{\phi_w}$ in full FP64 precision.
As we show in Section~\ref{sec:results} below,
one-body propagation can also be computed in mixed precision,
with little more loss in accuracy in the total energy
and modest additional speedup.
Our mixed-precision algorithm parallelizes over walkers even more efficiently than
AFQMC's typical embarrassingly parallel construction suggests,
as long as the set of walkers, trial wavefunction, and propagators
fit in the memory of one GPU.
Since the auxiliary field realizations differ,
each walker in the batch assigned to one GPU
requires its own distinct two-body propagator,
but they can be saved contiguously in memory and all the matrix multiplications
computed at once with the \textsc{cuBLAS} 12.9 \footnote{
\url{https://docs.nvidia.com/cuda/archive/12.9.0/cublas/index.html}}
routine \textsc{gemmStridedBatchedEx}
\footnote{See
\url{https://developer.nvidia.com/blog/cublas-strided-batched-matrix-multiply/}
and \url{https://docs.nvidia.com/cuda/archive/12.9.0/cublas/index.html\#cublasgemmstridedbatchedex};
the moniker \textsc{Ex} indicates that we can specify the precision.}.

We still sample the auxiliary fields, construct the two-body propagation
operator $\sqrt{-d\tau}\, \sum_\gamma x_\gamma \Op{L}_\gamma$, accumulate
$\ket{\phi_w}$, and compute the energy in full precision;
only the actual propagation step,
which for these systems constitutes the major part of the total computing time,
makes use of mixed-precision arithmetic.

\section{Results and Discussion} \label{sec:results}
\begin{figure}[ht]
    \includegraphics[width=\linewidth]{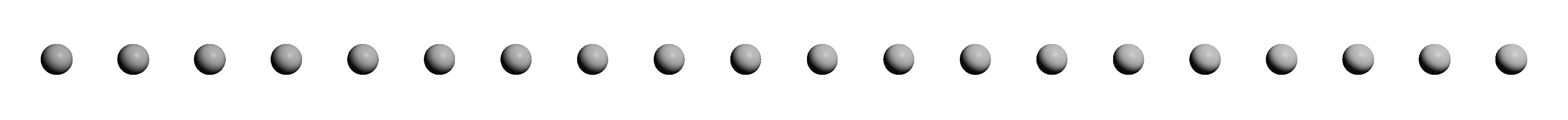}
    \caption{A short chain of 20 hydrogen atoms.
             We propagate chains up to \ce{H640} in the minimal STO-3G basis
             on a \qty{40}{\giga\byte} NVIDIA A100 GPU.
            }
    \label{fig:pretty-h-chain}
\end{figure}

We propagate one-dimensional chains of hydrogen atoms
for 200 AFQMC steps with $d\tau = \qty{0.005}{\au}$.
We use the minimal STO-3G basis~\cite{hehre_etal_1969},
with one contracted Gaussian-type orbital per hydrogen atom.
Figure~\ref{fig:pretty-h-chain} shows a chain of 20 atoms rendered with the
Raymol~\cite{castellanos_2026} fork of PyMOL~\cite{PyMOL}.

We use Los Alamos National Laboratory's Chicoma cluster
\cite{green_etal_2021},
which has 40 and \qty{80}{\giga\byte} A100 GPUs.
We implement our algorithm in a development version
of the \textsc{OpenMS} code
\footnote{A stable version of \textsc{OpenMS} is available at
\url{https://github.com/lanl/OpenMS}.}.
The initial state of the walkers is the restricted Hartree--Fock wavefunction
computed with \textsc{PySCF}~\cite{sun_etal_2018, sun_etal_2020, sun_etal_2026}.
Propagation on the CPU is accelerated with in-house \textsc{C++} code,
and GPU acceleration is achieved with the \textsc{CuPy} library \cite{okuta_etal_2017}.
For each comparison, we fix the same seed in the pseudorandom number generator
that realizes the values of $x_\gamma$.
Within each seed, every precision scheme therefore uses the same auxiliary-field samples.
For the purposes of timing, we warm up the propagation kernel
by running 10 warmup propagation steps before timing begins.

In the legends of the figures below, we use $\Op{T}$ as a shorthand for the
one-body propagation $e^{-d\tau\, \Op{T}/2}$ (which is applied twice per
propagation step) and $\Op{L}_\gamma$ as a shorthand for the two-body
propagation $e^{\sqrt{-d\tau}\, \sum_\gamma x_\gamma \Op{L}_\gamma}$.
We propagate the random walkers with five different methods:
\begin{enumerate}[itemsep=0.5em, label=(\arabic*)]
    \item Both the one- and two-body terms propagated in FP64
          (black line with round markers).
    \item One-body terms propagated in FP64, two-body terms in FP32
          (\emph{reduced}, not mixed, precision; green line with pentagonal markers).
    \item One- and two-body terms propagated in FP32 reduced precision
          (yellow line with square markers).
    \item One-body terms propagated with FP64, two-body terms in
          mixed-precision TF32 (Algorithm~\ref{alg:mult}; blue line with
          octagonal markers).
    \item One- and two-body terms propagated in mixed-precision TF32
          (purple line with triangular markers).
\end{enumerate}

\begin{figure}[htb]
    \centering
    \includegraphics[width=\linewidth]{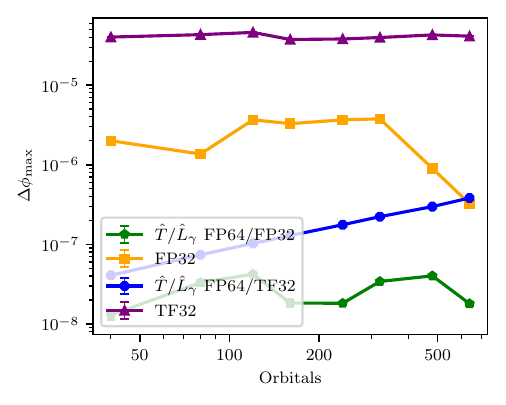}
    \caption{%
    The maximum relative error in the norm of the walker wavefunctions,
    $\Delta \phi_{\mathrm{max}} = 
    \max_{\tau}{\sum_w \lVert \phi_w^\tfp - \phi_w^\dfp \rVert / \lVert \phi_w^\dfp \rVert}$,
    for 100 walkers propagated over 200 AFQMC propagation steps
    and averaged over five random seeds.
    Unlike the local energy, the relative error in $\norm{\ket{\phi_w}}$
    does not increase monotonically with the system size.
    In particular, when both $\Op{T}$ and $\Op{L}_\gamma$
    are propagated with Algorithm \ref{alg:mult},
    the error saturates in the number of orbitals at \num{3e-5}.
    Nevertheless, using TF32 for $\Op{T}$ incurs more than an order of magnitude
    of additional error relative to propagating $\Op{T}$ in FP64.
    }
    \label{fig:err-phi-norb}
\end{figure}

In addition, we average GPU error metrics
(Figs.~\ref{fig:err-e-norb}--\ref{fig:err-e-nwalk}) over five random seeds.

Setting the \textsc{CuPy} environment variable \texttt{CUPY\_TF32}
forces single-precision (FP32) arithmetic to be routed through the Tensor Core
accelerators and performed in TF32 instead, albeit without the error mitigation
our mixed-precision algorithm provides.
As seen in the Supplementary Information,
propagation with this na\"ive TF32 arithmetic
is as fast as our mixed-precision scheme
(compare Fig.~\ref{fig:time-loglog} to Fig.~\ref{fig:supp-time}),
but substantially less accurate
(compare Figs.~\ref{fig:err-e-norb}--\ref{fig:err-e-nwalk}
against Figs.~\ref{fig:supp-norm}--\ref{fig:supp-err}).

\subsection{Efficiency}

\begin{figure}[htb]
    \centering
    \includegraphics[width=\linewidth]{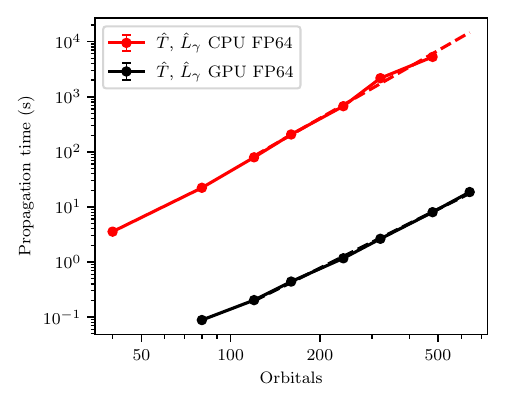}
    \caption{%
    The time required to propagate 100 walkers along 200 AFQMC steps
    in double (FP64) precision on the CPU (red) and the GPU (black),
    on a log-log scale.
    Error bars are the sample standard deviation after averaging five runs
    (in this case, all with the same random seed).
    The GPU is about two orders of magnitude faster than the CPU. 
    Note that the time to propagate 40 orbitals on the GPU
    is about the same as required to propagate 80,
    because the overhead of launching the GPU kernels
    is the bottleneck for small systems.
    }
    \label{fig:cpuvsgpu}
\end{figure}

Figure~\ref{fig:cpuvsgpu} compares FP64 propagation on the CPU and GPU.
The GPU is faster at every system size for which both measurements are available,
and the performance gap widens as the number of orbitals increases.
The CPU kernel takes more than \qty{5000}{\second}
to propagate 480 \ce{H} with 100 walkers atoms for 200 AFQMC steps;
the same propagation on the GPU takes \qty{8}{\second},
a speedup of 650x.
Note that FP64 propagation on the GPU
does not use our mixed-precision algorithm;
we simply route walkers and propagator matrices to the GPU and
replace the propagation routines with \textsc{CuPy} equivalents.

\begin{figure}[htb]
    \centering
    \includegraphics[width=\linewidth]{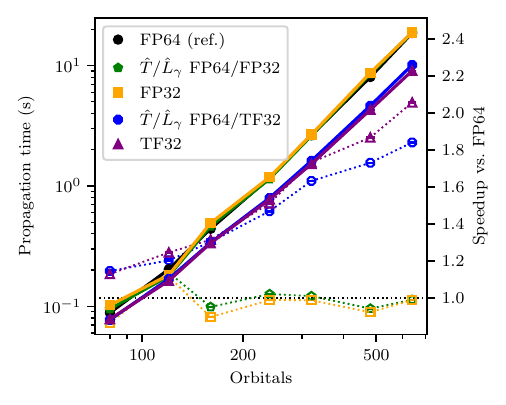}
    \caption{%
    Left axis, solid lines and markers:
    Wall time required to propagate 100 walkers along 200 AFQMC steps
    on the GPU in full, reduced, and mixed precision, plotted on a log-log scale.
    Right axis, dashed lines and open markers:
    Speedup relative to double precision.
    Both are averaged over the five random seeds.
    Note that we do not optimize the single-precision propagation,
    so on this GPU it is not much faster than double precision;
    mixed precision is faster in all systems.
    For 640 orbitals, propagating both one- and two-body terms in mixed precision
    is more than twice as fast as double-precision propagation.
    }
    \label{fig:time-loglog}
\end{figure}

We compare the five GPU propagation schemes in Figure~\ref{fig:time-loglog}.
The solid curves and filled markers represent the total time required to propagate
100 walkers along 200 AFQMC steps, shown on the left axis.
The dashed lines and open markers show the speedup relative to double-precision propagation on the right axis.
Single-precision (FP32) propagation, in yellow squares
(both one- and two-body terms propagated in FP32) and green pentagons
(one-body terms remain in FP64), is implemented na\"ively:
it calls the same routines as the FP64 GPU propagation, but with a
32-bit datatype (or its 64-bit complex equivalent).
It is included for comparing accuracy against our mixed-precision algorithm,
so we do not optimize it for speed, and, indeed, it is not noticeably faster than
FP64 propagation.
Our mixed-precision algorithm, in any case, is faster than FP64 or FP32
propagation for systems of moderate size.
For small systems, there is little practical benefit
to mixed-precision propagation,
as the computational kernels are bottlenecked by communication overhead;
by 120 orbitals, the speedup is about 20\%.
As with converting double-precision propagation to the GPU, however,
the benefit of propagating in mixed precision grows as the number of orbitals
increases. By 640 orbitals,
propagating both one- and two-body terms in mixed precision (purple triangles)
is twice as fast as using double precision.
This improvement, even relative to FP64 propagation on the GPU, 
is expected to continue until the size of $\{\Op{L}_\gamma\}$ exceeds
the memory available on the GPU.

\subsection{Accuracy}
\begin{figure}[htb]
    \centering
    \includegraphics[width=\linewidth]{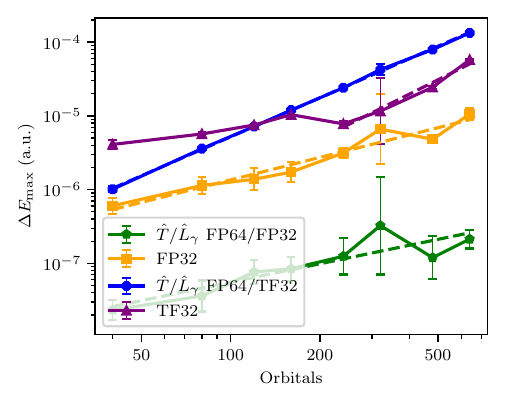}
    \caption{%
    The maximum value of the absolute error in the local energy, $\Delta E_{\max}$,
    in \si{\hartree}, of 100 walkers propagated over 200 AFQMC propagation steps
    and averaged over five random seeds.
    In larger systems, we observe some cancellation in the energy error
    when both one- and two-body terms are propagated in mixed-precision TF32 (purple line)
    compared to using double precision for the one-body propagator.
    However, the mixed-precision propagation has a less favorable scaling with
    system size. For 240 or more orbitals,
    the maximum error scales quadratically with the number of orbitals,
    while it is sub-quadratic or linear with other methods.
    Extrapolating each trendline suggests that the error could remain
    below \qty{1}{\milli\hartree} beyond 1500 orbitals.}
    \label{fig:err-e-norb}
\end{figure}

Propagating in reduced precision,
whether casting directly to FP32 or using Algorithm~\ref{alg:mult} in TF32,
incurs an accuracy penalty relative to FP64 propagation
greater than the FP32 machine epsilon $\eps_\sfp \approx 6 \times 10^{-8}$.
In Fig.~\ref{fig:err-e-norb}, we show the maximum error in the local energy
\begin{multline}
    E[\tau] = 
    \evo{\Op{H}_{\mc}} =
    \Big[ \frac{\sum_w c_w \mel{\Psi_\trial}{\Op{h}}{\phi_w}}
                {\sum_w c_w \braket{\Psi_\trial}{\phi_w}} \\ +
            \frac{\sum_w c_w \mel{\Psi_\trial}{\Op{V}_{\elec\elec}}{\phi_w}}
                 {\sum_w c_w \braket{\Psi_\trial}{\phi_w}} \Big] + C,
\end{multline}
of each propagation algorithm over the 200 propagation steps $\tau$,
relative to propagating in double precision.
We take the geometric mean of
$\Delta E_{\mathrm{max}} = \max_{\tau}{\abs{E[\tau] - E_{\dfp}[\tau]}}$
over five random seeds.
Propagating the two-body terms $\Op{L}_\gamma$ in single precision
while keeping the one-body term $\Op{T}$ in double precision
(green pentagons) incurs error little larger than $\eps_\sfp$.
This error increases sublinearly with system size.
Averaging over the five random seeds, the fitted line scales as $N^{0.83}$.
Propagating both one- and two-body terms in FP32 incurs a larger error,
which scales roughly linearly ($N^{1.01}$) with the number of orbitals
but remains below \qty{1e-5}{\hartree} at 640 orbitals,
the largest system we studied.
Mixed-precision arithmetic incurs a greater error than FP32 propagation.
If we propagate $\Op{L}_\gamma$ with Algorithm~\ref{alg:mult}
while keeping $\Op{T}$ in double precision (blue octagons), the error
scales as $N^{1.76}$ and exceeds \qty{1e-4}{\hartree} for 640 orbitals.
The error does not scale as a single power law when both $\Op{T}$
and $\Op{L}_\gamma$ are propagated in mixed precision (purple triangles).
For large systems, with 240 or more orbitals, the error scales as $N^{2.01}$, roughly quadratically.
However, we do see the potential for modest error cancellation
when the one- and two-body terms are both propagated in mixed precision.

This cancellation of error is not observed in the error in the
norm of the walker wavefunction,
\begin{equation}
    \Delta \phi_{\mathrm{max}}
        = \max_{\tau}{\sum_w \norm{\phi_w - \phi_{w,\dfp}}};
\end{equation}
$\Delta \phi_{\mathrm{max}}$ is typically 1--2 orders of magnitude greater
when both $\Op{L}_\gamma$ and $\Op{T}$ are propagated in mixed precision (purple triangles)
than when either $\Op{L}_\gamma$ alone (green pentagons)
or both $\Op{L}_\gamma$ and $\Op{T}$ (yellow squares) are propagated in single precision.
The dominant contribution instead comes from reducing the precision of $\Op{T}$.
Propagating $\Op{T}$ in FP64 lowers the error in the walker norm
by one to two orders of magnitude in both the FP32 and mixed-precision TF32 schemes.
We see a monotone increase in the error with system size only when
propagating $\Op{L}_\gamma$ in mixed precision and $\Op{T}$ in
double precision (blue octagons); otherwise, the error is nonmonotone
and indeed appears to saturate at roughly \num{3e-5} for
full mixed-precision propagation.
Note that, to allow exact comparison over more propagation steps,
we do not orthonormalize the walker wavefunctions in any of these schemes.
In a production-ready code, the walkers would be normalized every few propagation steps, which controls norm growth but does not, in general, make the reduced-precision and FP64 walkers identical.
The faster drift in the walker norms with our mixed-precision scheme is still worth noting.

\begin{figure}
    \centering
    \includegraphics[width=\linewidth]{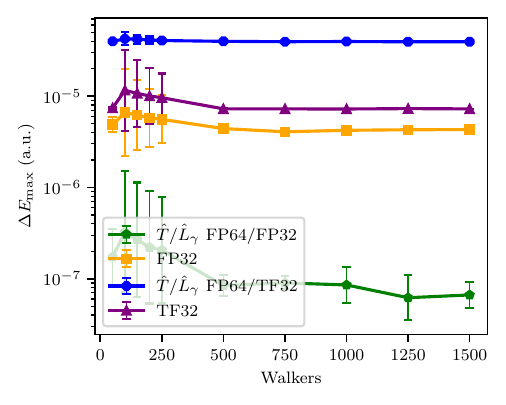}
    \caption{%
    Maximum error $\Delta E_{\mathrm{max}}$ in the local energy
    of 320 orbitals propagated for 200 AFQMC steps
    relative to FP64
    as a function of the number of walkers,
    averaged over five random seeds and on a logarithmic scale.
    Regardless of the propagation method, the error saturates
    in the number of walkers rather than increasing monotonically.
    Mixed precision incurs more error than single precision,
    but we observe modest error cancellation when propagating
    both one- and two-body terms in mixed precision
    (purple triangles) relative to propagating $\Op{T}$ in double precision
    (blue octagons).
    }
    \label{fig:err-e-nwalk}
\end{figure}

Finally, we observe in Fig.~\ref{fig:err-e-nwalk} that $\Delta E_{\mathrm{max}}$
does not increase with the number of random walkers when the number of orbitals
(320) and number of propagation steps (200) are held fixed.
The stochastic dependence on the random seed is, as expected,
larger for fewer walkers; the apparently larger error bars for propagating
$\Op{T}$ in double and $\Op{L}_\gamma$ in single precision (green pentagons)
is due to the logarithmic scale of $\Delta E_{\mathrm{max}}$.
For these seeds, we observe again the error cancellation
of Fig.~\ref{fig:err-e-norb} when both one- and two-body terms are propagated
with our mixed-precision algorithm
relative to full-precision propagation of $\Op{T}$.

\section{\label{sec:summary}Summary}
We have developed an efficient mixed-precision propagation scheme for AFQMC.
Utilizing NVIDIA GPUs' Tensor Cores,
which compute TF32-precision arithmetic extremely efficiently,
we substantially exceed the performance of propagating in either single (FP32)
or double (FP64) precision on the same GPU.
While our mixed-precision scheme suffers modest loss in the accuracy of the
local energy and does not maintain the norm of the walker wavefunctions
as well as propagating in single precision,
it remains well within the stochastic error inherent to AFQMC
and below the threshold of chemical accuracy for systems of moderate size.

The bottleneck for this method on GPUs is memory;
the entire set of Cholesky matrices $\{\Op{L}_\gamma\}$,
in addition to a collection of Slater determinant walker wavefunctions,
must fit into the main memory of the GPU.
Different walkers can be distributed to different GPUs as needed,
although the batched strided matrix multiplication available on Tensor Cores
means that the advantages of our mixed-precision algorithm are maximized
when as many walkers as possible are placed on one GPU.
However, the memory requirement for storing $\{\Op{L}_\gamma\}$
is $\Ord{(N_\AO^3)}$ since there are
$N_\gamma \sim \Ord{(N_\AO)}$ Cholesky matrices, each of size
$N_\AO \times N_\AO$~\cite{shi_zhang_2021}.
We note that the \textsc{ipie} software~\cite{malone_etal_2023}
alleviates the memory bottleneck
by structuring the Cholesky matrices hierarchically,
sharing a copy of $\{\Op{L}_\gamma\}$ among a small group of GPUs
instead of distributing it separately to each~\cite{jiang_etal_2024}.
In future work, we plan to investigate this and other schemes
for parallelizing $\{\Op{L}_\gamma\}$ over multiple GPUs.

%% file: kernel_acknow.tex
We acknowledge support from the U.S. Department of Energy,
Office of Science, Basic Energy Sciences,
Chemical Sciences, Geosciences, and Biosciences Division
under Triad National Security, LLC (``Triad'') contract Grant 89233218CNA000001
(FWP: LANLECF7).
This research used computational resources provided by
the Institutional Computing (IC) Program
at Los Alamos National Laboratory (LANL),
which is funded by
the Computational Systems and Software Environments subprogram
of LANL's Advanced Simulation and Computing program.
LANL is operated by Triad National Security, LLC,
for the National Nuclear Security Administration
of the U.S. Department of Energy (Contract No. 89233218CNA000001). 

%% file: kernel_supp.tex
\section{Na\"ive Tensor Core arithmetic}

\begin{figure}[ht]
    \centering
    \includegraphics[width=0.48\linewidth]{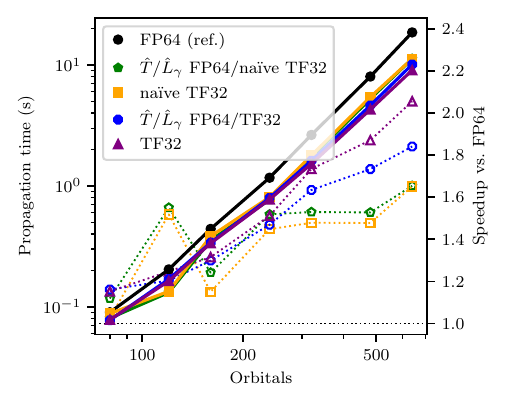}
    \caption{%
    The time for propagating 640 walkers 200 AFQMC steps on the GPU.
    Compare with Fig.~3 in the main text.}
    \label{fig:supp-time}
\end{figure}

On NVIDIA A100 machines, setting the environment variable
\texttt{CUPY\_TF32} to $1$ automatically routes single-precision arithmetic
through the Tensor Cores in TF32 precision. This na\"ive implementation is
almost as fast as our mixed-precision propagator on the systems tested
(Fig.~\ref{fig:supp-time}).

The mixed-precision algorithm still induces the largest error in the norm
$\Delta \phi_{\mathrm{max}} = \max_\tau \sum_w \norm{\phi_w - \phi_{w,\dfp}}$
of the walker wavefunctions. This could be because, when we split $\Op{L}_\gamma$ 
into high and low bits, there is more systematic drift relative even to setting the
low bits to zero.

\begin{figure}[ht]
    \centering
    \includegraphics[width=0.48\linewidth]{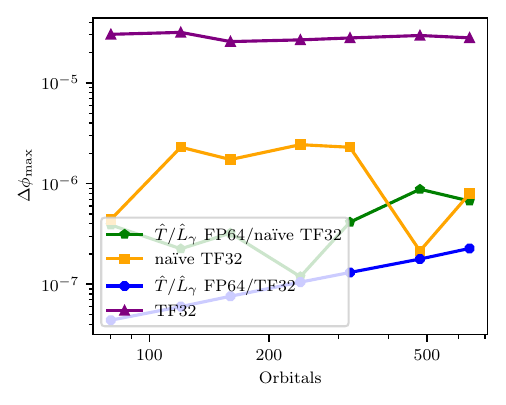}
    \caption{%
             The maximum error in the walker norm of 100 walkers over 200 propagation steps,
             as a function of the number of orbitals. Compared with Fig.~5 in the main text,
             propagating both $\Op{T}$ and $\Op{L}_\gamma$ in mixed precision (purple triangles)
             introduces more error than even the na\"ive TF32 propagation schemes
             (yellow squares, green pentagons).
             The dominant additional error is associated with reducing the precision of $\Op{T}$:
             as propagating $\Op{T}$ in double precision while using mixed precision for
             $\Op{L}_\gamma$ (blue octagons) has the lowest error relative to double-precision
             propagation.
             }
    \label{fig:supp-norm}
\end{figure}

\begin{figure*}[htb]
  \begin{tikzpicture}
    \node[inner sep=0] (imga) {\includegraphics[width=0.49\linewidth]{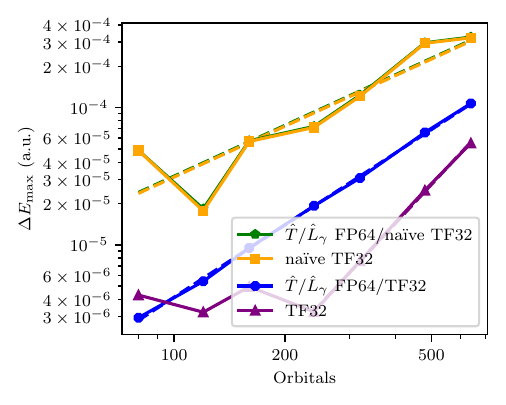}};
    \node[panel label, xshift=2pt, yshift=-5pt] at (imga.north west) {\small a)};
  \end{tikzpicture}
  \begin{tikzpicture}
    \node[inner sep=0] (imgb) {\includegraphics[width=0.49\linewidth]{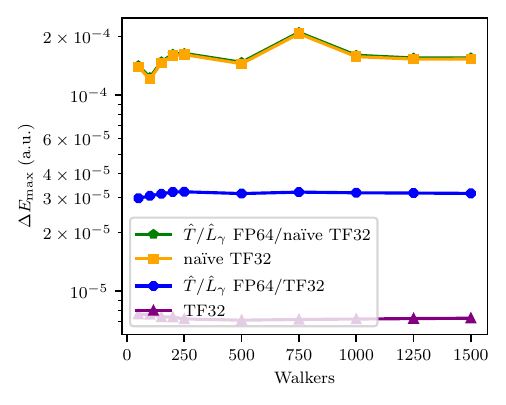}};
    \node[panel label, xshift=2pt, yshift=-5pt] at (imgb.north west) {\small b)};
  \end{tikzpicture}
    \caption{%
    Error in the local energy relative to FP64 propagation as a function of
    (a) the number of orbitals, (b) the number of walkers,
    with na\"ive TF32 arithmetic for $\Op{L}_\gamma$ (green pentagons) or for
    both $\Op{T}$ and $\Op{L}_\gamma$ (yellow squares).
    These runs are not averaged over different random seeds.
    Compare with Figs.~4--6 in the main text:
    na\"ive TF32 propagation (green pentagons, yellow squares)
    is substantially less accurate than mixed-precision propagation
    (blue octagons, purple triangles).
    }
    \label{fig:supp-err}
\end{figure*}

Na\"ive Tensor Core propagation does not ameliorate the reduced accuracy of TF32
relative to FP32 as our mixed-precision algorithm does.
Therefore, its error in the local energy is worse than our algorithm,
by roughly an order of magnitude up to 640 orbitals,
as seen in Fig.~\ref{fig:supp-err}.
Note, however, that $\Delta E_{\mathrm{max}}$ increases more steeply
with the number of walkers when both $\Op{T}$ and $\Op{L}_\gamma$
are propagated in mixed precision.

%% file: main.bbl
\begin{thebibliography}{78}%
\makeatletter
\providecommand \@ifxundefined [1]{%
 \@ifx{#1\undefined}
}%
\providecommand \@ifnum [1]{%
 \ifnum #1\expandafter \@firstoftwo
 \else \expandafter \@secondoftwo
 \fi
}%
\providecommand \@ifx [1]{%
 \ifx #1\expandafter \@firstoftwo
 \else \expandafter \@secondoftwo
 \fi
}%
\providecommand \natexlab [1]{#1}%
\providecommand \enquote  [1]{``#1''}%
\providecommand \bibnamefont  [1]{#1}%
\providecommand \bibfnamefont [1]{#1}%
\providecommand \citenamefont [1]{#1}%
\providecommand \href@noop [0]{\@secondoftwo}%
\providecommand \href [0]{\begingroup \@sanitize@url \@href}%
\providecommand \@href[1]{\@@startlink{#1}\@@href}%
\providecommand \@@href[1]{\endgroup#1\@@endlink}%
\providecommand \@sanitize@url [0]{\catcode `\\12\catcode `\$12\catcode `\&12\catcode `\#12\catcode `\^12\catcode `\_12\catcode `\%12\relax}%
\providecommand \@@startlink[1]{}%
\providecommand \@@endlink[0]{}%
\providecommand \url  [0]{\begingroup\@sanitize@url \@url }%
\providecommand \@url [1]{\endgroup\@href {#1}{\urlprefix }}%
\providecommand \urlprefix  [0]{URL }%
\providecommand \Eprint [0]{\href }%
\providecommand \doibase [0]{https://doi.org/}%
\providecommand \selectlanguage [0]{\@gobble}%
\providecommand \bibinfo  [0]{\@secondoftwo}%
\providecommand \bibfield  [0]{\@secondoftwo}%
\providecommand \translation [1]{[#1]}%
\providecommand \BibitemOpen [0]{}%
\providecommand \bibitemStop [0]{}%
\providecommand \bibitemNoStop [0]{.\EOS\space}%
\providecommand \EOS [0]{\spacefactor3000\relax}%
\providecommand \BibitemShut  [1]{\csname bibitem#1\endcsname}%
\let\auto@bib@innerbib\@empty
\bibitem [{\citenamefont {Greeley}\ \emph {et~al.}(2002)\citenamefont {Greeley}, \citenamefont {N{\o}rskov},\ and\ \citenamefont {Mavrikakis}}]{greeley_etal_2002}%
  \BibitemOpen
  \bibfield  {author} {\bibinfo {author} {\bibfnamefont {J.}~\bibnamefont {Greeley}}, \bibinfo {author} {\bibfnamefont {J.~K.}\ \bibnamefont {N{\o}rskov}},\ and\ \bibinfo {author} {\bibfnamefont {M.}~\bibnamefont {Mavrikakis}},\ }\href {https://doi.org/10.1146/annurev.physchem.53.100301.131630} {\bibfield  {journal} {\bibinfo  {journal} {Annual Review of Physical Chemistry}\ }\textbf {\bibinfo {volume} {53}},\ \bibinfo {pages} {319} (\bibinfo {year} {2002})}\BibitemShut {NoStop}%
\bibitem [{\citenamefont {Marzari}\ \emph {et~al.}(2021)\citenamefont {Marzari}, \citenamefont {Ferretti},\ and\ \citenamefont {Wolverton}}]{marzari_etal_2021}%
  \BibitemOpen
  \bibfield  {author} {\bibinfo {author} {\bibfnamefont {N.}~\bibnamefont {Marzari}}, \bibinfo {author} {\bibfnamefont {A.}~\bibnamefont {Ferretti}},\ and\ \bibinfo {author} {\bibfnamefont {C.}~\bibnamefont {Wolverton}},\ }\href {https://doi.org/10.1038/s41563-021-01013-3} {\bibfield  {journal} {\bibinfo  {journal} {Nature Materials}\ }\textbf {\bibinfo {volume} {20}},\ \bibinfo {pages} {736} (\bibinfo {year} {2021})}\BibitemShut {NoStop}%
\bibitem [{\citenamefont {Van~Noorden}(2025)}]{vannoorden_2025}%
  \BibitemOpen
  \bibfield  {author} {\bibinfo {author} {\bibfnamefont {R.}~\bibnamefont {Van~Noorden}},\ }\href {https://doi.org/10.1038/d41586-025-01124-w} {\bibfield  {journal} {\bibinfo  {journal} {Nature}\ }\textbf {\bibinfo {volume} {640}},\ \bibinfo {pages} {591} (\bibinfo {year} {2025})}\BibitemShut {NoStop}%
\bibitem [{\citenamefont {Lee}\ \emph {et~al.}(1988)\citenamefont {Lee}, \citenamefont {Yang},\ and\ \citenamefont {Parr}}]{lee_etal_1988}%
  \BibitemOpen
  \bibfield  {author} {\bibinfo {author} {\bibfnamefont {C.}~\bibnamefont {Lee}}, \bibinfo {author} {\bibfnamefont {W.}~\bibnamefont {Yang}},\ and\ \bibinfo {author} {\bibfnamefont {R.~G.}\ \bibnamefont {Parr}},\ }\href {https://doi.org/10.1103/PhysRevB.37.785} {\bibfield  {journal} {\bibinfo  {journal} {Physical Review B}\ }\textbf {\bibinfo {volume} {37}},\ \bibinfo {pages} {785} (\bibinfo {year} {1988})}\BibitemShut {NoStop}%
\bibitem [{\citenamefont {Perdew}\ \emph {et~al.}(1996)\citenamefont {Perdew}, \citenamefont {Burke},\ and\ \citenamefont {Ernzerhof}}]{perdew_etal_1996a}%
  \BibitemOpen
  \bibfield  {author} {\bibinfo {author} {\bibfnamefont {J.~P.}\ \bibnamefont {Perdew}}, \bibinfo {author} {\bibfnamefont {K.}~\bibnamefont {Burke}},\ and\ \bibinfo {author} {\bibfnamefont {M.}~\bibnamefont {Ernzerhof}},\ }\href {https://doi.org/10.1103/PhysRevLett.77.3865} {\bibfield  {journal} {\bibinfo  {journal} {Physical Review Letters}\ }\textbf {\bibinfo {volume} {77}},\ \bibinfo {pages} {3865} (\bibinfo {year} {1996})}\BibitemShut {NoStop}%
\bibitem [{\citenamefont {Kresse}\ and\ \citenamefont {Furthm{\"u}ller}(1996)}]{kresse_furthmuller_1996}%
  \BibitemOpen
  \bibfield  {author} {\bibinfo {author} {\bibfnamefont {G.}~\bibnamefont {Kresse}}\ and\ \bibinfo {author} {\bibfnamefont {J.}~\bibnamefont {Furthm{\"u}ller}},\ }\href {https://doi.org/10.1103/PhysRevB.54.11169} {\bibfield  {journal} {\bibinfo  {journal} {Physical Review B}\ }\textbf {\bibinfo {volume} {54}},\ \bibinfo {pages} {11169} (\bibinfo {year} {1996})}\BibitemShut {NoStop}%
\bibitem [{\citenamefont {Teale}\ \emph {et~al.}(2022)\citenamefont {Teale}, \citenamefont {Helgaker}, \citenamefont {Savin}, \citenamefont {Adamo}, \citenamefont {Aradi}, \citenamefont {Arbuznikov}, \citenamefont {Ayers}, \citenamefont {Baerends}, \citenamefont {Barone}, \citenamefont {Calaminici}, \citenamefont {Canc{\`e}s}, \citenamefont {Carter}, \citenamefont {Chattaraj}, \citenamefont {Chermette}, \citenamefont {Ciofini}, \citenamefont {Crawford}, \citenamefont {De~Proft}, \citenamefont {Dobson}, \citenamefont {Draxl}, \citenamefont {Frauenheim}, \citenamefont {Fromager}, \citenamefont {Fuentealba}, \citenamefont {Gagliardi}, \citenamefont {Galli}, \citenamefont {Gao}, \citenamefont {Geerlings}, \citenamefont {Gidopoulos}, \citenamefont {Gill}, \citenamefont {{Gori-Giorgi}}, \citenamefont {G{\"o}rling}, \citenamefont {Gould}, \citenamefont {Grimme}, \citenamefont {Gritsenko}, \citenamefont {Jensen}, \citenamefont {Johnson}, \citenamefont {Jones}, \citenamefont {Kaupp}, \citenamefont {K{\"o}ster},
  \citenamefont {Kronik}, \citenamefont {Krylov}, \citenamefont {Kvaal}, \citenamefont {Laestadius}, \citenamefont {Levy}, \citenamefont {Lewin}, \citenamefont {Liu}, \citenamefont {Loos}, \citenamefont {Maitra}, \citenamefont {Neese}, \citenamefont {Perdew}, \citenamefont {Pernal}, \citenamefont {Pernot}, \citenamefont {Piecuch}, \citenamefont {Rebolini}, \citenamefont {Reining}, \citenamefont {Romaniello}, \citenamefont {Ruzsinszky}, \citenamefont {Salahub}, \citenamefont {Scheffler}, \citenamefont {Schwerdtfeger}, \citenamefont {Staroverov}, \citenamefont {Sun}, \citenamefont {Tellgren}, \citenamefont {Tozer}, \citenamefont {Trickey}, \citenamefont {Ullrich}, \citenamefont {Vela}, \citenamefont {Vignale}, \citenamefont {Wesolowski}, \citenamefont {Xu},\ and\ \citenamefont {Yang}}]{teale_etal_2022}%
  \BibitemOpen
  \bibfield  {author} {\bibinfo {author} {\bibfnamefont {A.~M.}\ \bibnamefont {Teale}}, \bibinfo {author} {\bibfnamefont {T.}~\bibnamefont {Helgaker}}, \bibinfo {author} {\bibfnamefont {A.}~\bibnamefont {Savin}}, \bibinfo {author} {\bibfnamefont {C.}~\bibnamefont {Adamo}}, \bibinfo {author} {\bibfnamefont {B.}~\bibnamefont {Aradi}}, \bibinfo {author} {\bibfnamefont {A.~V.}\ \bibnamefont {Arbuznikov}}, \bibinfo {author} {\bibfnamefont {P.~W.}\ \bibnamefont {Ayers}}, \bibinfo {author} {\bibfnamefont {E.~J.}\ \bibnamefont {Baerends}}, \bibinfo {author} {\bibfnamefont {V.}~\bibnamefont {Barone}}, \bibinfo {author} {\bibfnamefont {P.}~\bibnamefont {Calaminici}}, \bibinfo {author} {\bibfnamefont {E.}~\bibnamefont {Canc{\`e}s}}, \bibinfo {author} {\bibfnamefont {E.~A.}\ \bibnamefont {Carter}}, \bibinfo {author} {\bibfnamefont {P.~K.}\ \bibnamefont {Chattaraj}}, \bibinfo {author} {\bibfnamefont {H.}~\bibnamefont {Chermette}}, \bibinfo {author} {\bibfnamefont {I.}~\bibnamefont {Ciofini}}, \bibinfo {author}
  {\bibfnamefont {T.~D.}\ \bibnamefont {Crawford}}, \bibinfo {author} {\bibfnamefont {F.}~\bibnamefont {De~Proft}}, \bibinfo {author} {\bibfnamefont {J.~F.}\ \bibnamefont {Dobson}}, \bibinfo {author} {\bibfnamefont {C.}~\bibnamefont {Draxl}}, \bibinfo {author} {\bibfnamefont {T.}~\bibnamefont {Frauenheim}}, \bibinfo {author} {\bibfnamefont {E.}~\bibnamefont {Fromager}}, \bibinfo {author} {\bibfnamefont {P.}~\bibnamefont {Fuentealba}}, \bibinfo {author} {\bibfnamefont {L.}~\bibnamefont {Gagliardi}}, \bibinfo {author} {\bibfnamefont {G.}~\bibnamefont {Galli}}, \bibinfo {author} {\bibfnamefont {J.}~\bibnamefont {Gao}}, \bibinfo {author} {\bibfnamefont {P.}~\bibnamefont {Geerlings}}, \bibinfo {author} {\bibfnamefont {N.}~\bibnamefont {Gidopoulos}}, \bibinfo {author} {\bibfnamefont {P.~M.~W.}\ \bibnamefont {Gill}}, \bibinfo {author} {\bibfnamefont {P.}~\bibnamefont {{Gori-Giorgi}}}, \bibinfo {author} {\bibfnamefont {A.}~\bibnamefont {G{\"o}rling}}, \bibinfo {author} {\bibfnamefont {T.}~\bibnamefont {Gould}},
  \bibinfo {author} {\bibfnamefont {S.}~\bibnamefont {Grimme}}, \bibinfo {author} {\bibfnamefont {O.}~\bibnamefont {Gritsenko}}, \bibinfo {author} {\bibfnamefont {H.~J.~A.}\ \bibnamefont {Jensen}}, \bibinfo {author} {\bibfnamefont {E.~R.}\ \bibnamefont {Johnson}}, \bibinfo {author} {\bibfnamefont {R.~O.}\ \bibnamefont {Jones}}, \bibinfo {author} {\bibfnamefont {M.}~\bibnamefont {Kaupp}}, \bibinfo {author} {\bibfnamefont {A.~M.}\ \bibnamefont {K{\"o}ster}}, \bibinfo {author} {\bibfnamefont {L.}~\bibnamefont {Kronik}}, \bibinfo {author} {\bibfnamefont {A.~I.}\ \bibnamefont {Krylov}}, \bibinfo {author} {\bibfnamefont {S.}~\bibnamefont {Kvaal}}, \bibinfo {author} {\bibfnamefont {A.}~\bibnamefont {Laestadius}}, \bibinfo {author} {\bibfnamefont {M.}~\bibnamefont {Levy}}, \bibinfo {author} {\bibfnamefont {M.}~\bibnamefont {Lewin}}, \bibinfo {author} {\bibfnamefont {S.}~\bibnamefont {Liu}}, \bibinfo {author} {\bibfnamefont {P.-F.}\ \bibnamefont {Loos}}, \bibinfo {author} {\bibfnamefont {N.~T.}\ \bibnamefont
  {Maitra}}, \bibinfo {author} {\bibfnamefont {F.}~\bibnamefont {Neese}}, \bibinfo {author} {\bibfnamefont {J.~P.}\ \bibnamefont {Perdew}}, \bibinfo {author} {\bibfnamefont {K.}~\bibnamefont {Pernal}}, \bibinfo {author} {\bibfnamefont {P.}~\bibnamefont {Pernot}}, \bibinfo {author} {\bibfnamefont {P.}~\bibnamefont {Piecuch}}, \bibinfo {author} {\bibfnamefont {E.}~\bibnamefont {Rebolini}}, \bibinfo {author} {\bibfnamefont {L.}~\bibnamefont {Reining}}, \bibinfo {author} {\bibfnamefont {P.}~\bibnamefont {Romaniello}}, \bibinfo {author} {\bibfnamefont {A.}~\bibnamefont {Ruzsinszky}}, \bibinfo {author} {\bibfnamefont {D.~R.}\ \bibnamefont {Salahub}}, \bibinfo {author} {\bibfnamefont {M.}~\bibnamefont {Scheffler}}, \bibinfo {author} {\bibfnamefont {P.}~\bibnamefont {Schwerdtfeger}}, \bibinfo {author} {\bibfnamefont {V.~N.}\ \bibnamefont {Staroverov}}, \bibinfo {author} {\bibfnamefont {J.}~\bibnamefont {Sun}}, \bibinfo {author} {\bibfnamefont {E.}~\bibnamefont {Tellgren}}, \bibinfo {author} {\bibfnamefont {D.~J.}\
  \bibnamefont {Tozer}}, \bibinfo {author} {\bibfnamefont {S.~B.}\ \bibnamefont {Trickey}}, \bibinfo {author} {\bibfnamefont {C.~A.}\ \bibnamefont {Ullrich}}, \bibinfo {author} {\bibfnamefont {A.}~\bibnamefont {Vela}}, \bibinfo {author} {\bibfnamefont {G.}~\bibnamefont {Vignale}}, \bibinfo {author} {\bibfnamefont {T.~A.}\ \bibnamefont {Wesolowski}}, \bibinfo {author} {\bibfnamefont {X.}~\bibnamefont {Xu}},\ and\ \bibinfo {author} {\bibfnamefont {W.}~\bibnamefont {Yang}},\ }\href {https://doi.org/10.1039/d2cp02827a} {\bibfield  {journal} {\bibinfo  {journal} {Physical Chemistry Chemical Physics}\ }\textbf {\bibinfo {volume} {24}},\ \bibinfo {pages} {28700} (\bibinfo {year} {2022})}\BibitemShut {NoStop}%
\bibitem [{\citenamefont {Neese}(2006)}]{neese_2006}%
  \BibitemOpen
  \bibfield  {author} {\bibinfo {author} {\bibfnamefont {F.}~\bibnamefont {Neese}},\ }\href {https://doi.org/10.1007/s00775-006-0138-1} {\bibfield  {journal} {\bibinfo  {journal} {JBIC Journal of Biological Inorganic Chemistry}\ }\textbf {\bibinfo {volume} {11}},\ \bibinfo {pages} {702} (\bibinfo {year} {2006})}\BibitemShut {NoStop}%
\bibitem [{\citenamefont {Pavarini}(2021)}]{pavarini_2021}%
  \BibitemOpen
  \bibfield  {author} {\bibinfo {author} {\bibfnamefont {E.}~\bibnamefont {Pavarini}},\ }\href {https://doi.org/10.1007/s40766-021-00025-8} {\bibfield  {journal} {\bibinfo  {journal} {La Rivista del Nuovo Cimento}\ }\textbf {\bibinfo {volume} {44}},\ \bibinfo {pages} {597} (\bibinfo {year} {2021})}\BibitemShut {NoStop}%
\bibitem [{\citenamefont {Gavini}\ \emph {et~al.}(2023)\citenamefont {Gavini}, \citenamefont {Baroni}, \citenamefont {Blum}, \citenamefont {Bowler}, \citenamefont {Buccheri}, \citenamefont {Chelikowsky}, \citenamefont {Das}, \citenamefont {Dawson}, \citenamefont {Delugas}, \citenamefont {Dogan}, \citenamefont {Draxl}, \citenamefont {Galli}, \citenamefont {Genovese}, \citenamefont {Giannozzi}, \citenamefont {Giantomassi}, \citenamefont {Gonze}, \citenamefont {Govoni}, \citenamefont {Gygi}, \citenamefont {Gulans}, \citenamefont {Herbert}, \citenamefont {Kokott}, \citenamefont {K{\"u}hne}, \citenamefont {Liou}, \citenamefont {Miyazaki}, \citenamefont {Motamarri}, \citenamefont {Nakata}, \citenamefont {Pask}, \citenamefont {Plessl}, \citenamefont {Ratcliff}, \citenamefont {Richard}, \citenamefont {Rossi}, \citenamefont {Schade}, \citenamefont {Scheffler}, \citenamefont {Sch{\"u}tt}, \citenamefont {Suryanarayana}, \citenamefont {Torrent}, \citenamefont {Truflandier}, \citenamefont {Windus}, \citenamefont {Xu},
  \citenamefont {Yu},\ and\ \citenamefont {Perez}}]{gavini_etal_2023}%
  \BibitemOpen
  \bibfield  {author} {\bibinfo {author} {\bibfnamefont {V.}~\bibnamefont {Gavini}}, \bibinfo {author} {\bibfnamefont {S.}~\bibnamefont {Baroni}}, \bibinfo {author} {\bibfnamefont {V.}~\bibnamefont {Blum}}, \bibinfo {author} {\bibfnamefont {D.~R.}\ \bibnamefont {Bowler}}, \bibinfo {author} {\bibfnamefont {A.}~\bibnamefont {Buccheri}}, \bibinfo {author} {\bibfnamefont {J.~R.}\ \bibnamefont {Chelikowsky}}, \bibinfo {author} {\bibfnamefont {S.}~\bibnamefont {Das}}, \bibinfo {author} {\bibfnamefont {W.}~\bibnamefont {Dawson}}, \bibinfo {author} {\bibfnamefont {P.}~\bibnamefont {Delugas}}, \bibinfo {author} {\bibfnamefont {M.}~\bibnamefont {Dogan}}, \bibinfo {author} {\bibfnamefont {C.}~\bibnamefont {Draxl}}, \bibinfo {author} {\bibfnamefont {G.}~\bibnamefont {Galli}}, \bibinfo {author} {\bibfnamefont {L.}~\bibnamefont {Genovese}}, \bibinfo {author} {\bibfnamefont {P.}~\bibnamefont {Giannozzi}}, \bibinfo {author} {\bibfnamefont {M.}~\bibnamefont {Giantomassi}}, \bibinfo {author} {\bibfnamefont {X.}~\bibnamefont
  {Gonze}}, \bibinfo {author} {\bibfnamefont {M.}~\bibnamefont {Govoni}}, \bibinfo {author} {\bibfnamefont {F.}~\bibnamefont {Gygi}}, \bibinfo {author} {\bibfnamefont {A.}~\bibnamefont {Gulans}}, \bibinfo {author} {\bibfnamefont {J.~M.}\ \bibnamefont {Herbert}}, \bibinfo {author} {\bibfnamefont {S.}~\bibnamefont {Kokott}}, \bibinfo {author} {\bibfnamefont {T.~D.}\ \bibnamefont {K{\"u}hne}}, \bibinfo {author} {\bibfnamefont {K.-H.}\ \bibnamefont {Liou}}, \bibinfo {author} {\bibfnamefont {T.}~\bibnamefont {Miyazaki}}, \bibinfo {author} {\bibfnamefont {P.}~\bibnamefont {Motamarri}}, \bibinfo {author} {\bibfnamefont {A.}~\bibnamefont {Nakata}}, \bibinfo {author} {\bibfnamefont {J.~E.}\ \bibnamefont {Pask}}, \bibinfo {author} {\bibfnamefont {C.}~\bibnamefont {Plessl}}, \bibinfo {author} {\bibfnamefont {L.~E.}\ \bibnamefont {Ratcliff}}, \bibinfo {author} {\bibfnamefont {R.~M.}\ \bibnamefont {Richard}}, \bibinfo {author} {\bibfnamefont {M.}~\bibnamefont {Rossi}}, \bibinfo {author} {\bibfnamefont {R.}~\bibnamefont
  {Schade}}, \bibinfo {author} {\bibfnamefont {M.}~\bibnamefont {Scheffler}}, \bibinfo {author} {\bibfnamefont {O.}~\bibnamefont {Sch{\"u}tt}}, \bibinfo {author} {\bibfnamefont {P.}~\bibnamefont {Suryanarayana}}, \bibinfo {author} {\bibfnamefont {M.}~\bibnamefont {Torrent}}, \bibinfo {author} {\bibfnamefont {L.}~\bibnamefont {Truflandier}}, \bibinfo {author} {\bibfnamefont {T.~L.}\ \bibnamefont {Windus}}, \bibinfo {author} {\bibfnamefont {Q.}~\bibnamefont {Xu}}, \bibinfo {author} {\bibfnamefont {V.~W.-Z.}\ \bibnamefont {Yu}},\ and\ \bibinfo {author} {\bibfnamefont {D.}~\bibnamefont {Perez}},\ }\href {https://doi.org/10.1088/1361-651X/acdf06} {\bibfield  {journal} {\bibinfo  {journal} {Modelling and Simulation in Materials Science and Engineering}\ }\textbf {\bibinfo {volume} {31}},\ \bibinfo {pages} {063301} (\bibinfo {year} {2023})}\BibitemShut {NoStop}%
\bibitem [{\citenamefont {Bianchini}\ \emph {et~al.}(2026)\citenamefont {Bianchini}, \citenamefont {Geuna},\ and\ \citenamefont {Shermatov}}]{bianchini_etal_2026}%
  \BibitemOpen
  \bibfield  {author} {\bibinfo {author} {\bibfnamefont {S.}~\bibnamefont {Bianchini}}, \bibinfo {author} {\bibfnamefont {A.}~\bibnamefont {Geuna}},\ and\ \bibinfo {author} {\bibfnamefont {F.}~\bibnamefont {Shermatov}},\ }\bibfield  {journal} {\bibinfo  {journal} {Scientific Reports}\ }\href {https://doi.org/10.1038/s41598-026-63438-7} {10.1038/s41598-026-63438-7} (\bibinfo {year} {2026})\BibitemShut {NoStop}%
\bibitem [{\citenamefont {Resch}\ \emph {et~al.}(2025)\citenamefont {Resch}, \citenamefont {Gebert},\ and\ \citenamefont {Hoppe}}]{resch_etal_2025}%
  \BibitemOpen
  \bibfield  {author} {\bibinfo {author} {\bibfnamefont {M.~M.}\ \bibnamefont {Resch}}, \bibinfo {author} {\bibfnamefont {J.}~\bibnamefont {Gebert}},\ and\ \bibinfo {author} {\bibfnamefont {D.}~\bibnamefont {Hoppe}},\ }in\ \href {https://doi.org/10.1109/IC363308.2025.10956516} {\emph {\bibinfo {booktitle} {2025 {{International Conference}} on {{Intelligent Control}}, {{Computing}} and {{Communications}} ({{IC3}})}}}\ (\bibinfo {year} {2025})\ pp.\ \bibinfo {pages} {897--904}\BibitemShut {NoStop}%
\bibitem [{\citenamefont {Gupta}\ \emph {et~al.}(2015)\citenamefont {Gupta}, \citenamefont {Agrawal}, \citenamefont {Gopalakrishnan},\ and\ \citenamefont {Narayanan}}]{gupta_etal_2015}%
  \BibitemOpen
  \bibfield  {author} {\bibinfo {author} {\bibfnamefont {S.}~\bibnamefont {Gupta}}, \bibinfo {author} {\bibfnamefont {A.}~\bibnamefont {Agrawal}}, \bibinfo {author} {\bibfnamefont {K.}~\bibnamefont {Gopalakrishnan}},\ and\ \bibinfo {author} {\bibfnamefont {P.}~\bibnamefont {Narayanan}},\ }in\ \href@noop {} {\emph {\bibinfo {booktitle} {Proceedings of the 32nd {{International Conference}} on {{Machine Learning}}}}}\ (\bibinfo  {publisher} {PMLR},\ \bibinfo {year} {2015})\ pp.\ \bibinfo {pages} {1737--1746}\BibitemShut {NoStop}%
\bibitem [{\citenamefont {Dean}(2020)}]{dean_2020}%
  \BibitemOpen
  \bibfield  {author} {\bibinfo {author} {\bibfnamefont {J.}~\bibnamefont {Dean}},\ }in\ \href {https://doi.org/10.1109/ISSCC19947.2020.9063049} {\emph {\bibinfo {booktitle} {2020 {{IEEE International Solid-State Circuits Conference}} - ({{ISSCC}})}}}\ (\bibinfo {year} {2020})\ pp.\ \bibinfo {pages} {8--14}\BibitemShut {NoStop}%
\bibitem [{\citenamefont {Shinde}\ \emph {et~al.}(2025)\citenamefont {Shinde}, \citenamefont {Filippi}, \citenamefont {Scemama},\ and\ \citenamefont {Jalby}}]{shinde_etal_2025a}%
  \BibitemOpen
  \bibfield  {author} {\bibinfo {author} {\bibfnamefont {R.}~\bibnamefont {Shinde}}, \bibinfo {author} {\bibfnamefont {C.}~\bibnamefont {Filippi}}, \bibinfo {author} {\bibfnamefont {A.}~\bibnamefont {Scemama}},\ and\ \bibinfo {author} {\bibfnamefont {W.}~\bibnamefont {Jalby}},\ }\href {https://doi.org/10.1038/s42254-025-00823-7} {\bibfield  {journal} {\bibinfo  {journal} {Nature Reviews Physics}\ }\textbf {\bibinfo {volume} {7}},\ \bibinfo {pages} {378} (\bibinfo {year} {2025})}\BibitemShut {NoStop}%
\bibitem [{\citenamefont {Shi}\ and\ \citenamefont {Zhang}(2021)}]{shi_zhang_2021}%
  \BibitemOpen
  \bibfield  {author} {\bibinfo {author} {\bibfnamefont {H.}~\bibnamefont {Shi}}\ and\ \bibinfo {author} {\bibfnamefont {S.}~\bibnamefont {Zhang}},\ }\href {https://doi.org/10.1063/5.0031024} {\bibfield  {journal} {\bibinfo  {journal} {The Journal of Chemical Physics}\ }\textbf {\bibinfo {volume} {154}},\ \bibinfo {pages} {024107} (\bibinfo {year} {2021})}\BibitemShut {NoStop}%
\bibitem [{\citenamefont {Lee}\ \emph {et~al.}(2022)\citenamefont {Lee}, \citenamefont {Pham},\ and\ \citenamefont {Reichman}}]{lee_etal_2022}%
  \BibitemOpen
  \bibfield  {author} {\bibinfo {author} {\bibfnamefont {J.}~\bibnamefont {Lee}}, \bibinfo {author} {\bibfnamefont {H.~Q.}\ \bibnamefont {Pham}},\ and\ \bibinfo {author} {\bibfnamefont {D.~R.}\ \bibnamefont {Reichman}},\ }\href {https://doi.org/10.1021/acs.jctc.2c00802} {\bibfield  {journal} {\bibinfo  {journal} {Journal of Chemical Theory and Computation}\ }\textbf {\bibinfo {volume} {18}},\ \bibinfo {pages} {7024} (\bibinfo {year} {2022})}\BibitemShut {NoStop}%
\bibitem [{\citenamefont {Shee}\ \emph {et~al.}(2023)\citenamefont {Shee}, \citenamefont {Weber}, \citenamefont {Reichman}, \citenamefont {Friesner},\ and\ \citenamefont {Zhang}}]{shee_etal_2023}%
  \BibitemOpen
  \bibfield  {author} {\bibinfo {author} {\bibfnamefont {J.}~\bibnamefont {Shee}}, \bibinfo {author} {\bibfnamefont {J.~L.}\ \bibnamefont {Weber}}, \bibinfo {author} {\bibfnamefont {D.~R.}\ \bibnamefont {Reichman}}, \bibinfo {author} {\bibfnamefont {R.~A.}\ \bibnamefont {Friesner}},\ and\ \bibinfo {author} {\bibfnamefont {S.}~\bibnamefont {Zhang}},\ }\href {https://doi.org/10.1063/5.0134009} {\bibfield  {journal} {\bibinfo  {journal} {The Journal of Chemical Physics}\ }\textbf {\bibinfo {volume} {158}},\ \bibinfo {pages} {140901} (\bibinfo {year} {2023})}\BibitemShut {NoStop}%
\bibitem [{\citenamefont {Church}\ and\ \citenamefont {Rubenstein}(2021)}]{church_rubenstein_2021}%
  \BibitemOpen
  \bibfield  {author} {\bibinfo {author} {\bibfnamefont {M.~S.}\ \bibnamefont {Church}}\ and\ \bibinfo {author} {\bibfnamefont {B.~M.}\ \bibnamefont {Rubenstein}},\ }\href {https://doi.org/10.1063/5.0049116} {\bibfield  {journal} {\bibinfo  {journal} {The Journal of Chemical Physics}\ }\textbf {\bibinfo {volume} {154}},\ \bibinfo {pages} {184103} (\bibinfo {year} {2021})}\BibitemShut {NoStop}%
\bibitem [{\citenamefont {Shee}\ \emph {et~al.}(2018)\citenamefont {Shee}, \citenamefont {Arthur}, \citenamefont {Zhang}, \citenamefont {Reichman},\ and\ \citenamefont {Friesner}}]{shee_etal_2018}%
  \BibitemOpen
  \bibfield  {author} {\bibinfo {author} {\bibfnamefont {J.}~\bibnamefont {Shee}}, \bibinfo {author} {\bibfnamefont {E.~J.}\ \bibnamefont {Arthur}}, \bibinfo {author} {\bibfnamefont {S.}~\bibnamefont {Zhang}}, \bibinfo {author} {\bibfnamefont {D.~R.}\ \bibnamefont {Reichman}},\ and\ \bibinfo {author} {\bibfnamefont {R.~A.}\ \bibnamefont {Friesner}},\ }\href {https://doi.org/10.1021/acs.jctc.8b00342} {\bibfield  {journal} {\bibinfo  {journal} {Journal of Chemical Theory and Computation}\ }\textbf {\bibinfo {volume} {14}},\ \bibinfo {pages} {4109} (\bibinfo {year} {2018})}\BibitemShut {NoStop}%
\bibitem [{\citenamefont {R{\o}eggen}\ and\ \citenamefont {{Wisl{\o}ff-Nilssen}}(1986)}]{roeggen_wisloff-nilssen_1986}%
  \BibitemOpen
  \bibfield  {author} {\bibinfo {author} {\bibfnamefont {I.}~\bibnamefont {R{\o}eggen}}\ and\ \bibinfo {author} {\bibfnamefont {E.}~\bibnamefont {{Wisl{\o}ff-Nilssen}}},\ }\href {https://doi.org/10.1016/0009-2614(86)80099-9} {\bibfield  {journal} {\bibinfo  {journal} {Chemical Physics Letters}\ }\textbf {\bibinfo {volume} {132}},\ \bibinfo {pages} {154} (\bibinfo {year} {1986})}\BibitemShut {NoStop}%
\bibitem [{\citenamefont {Pedersen}\ \emph {et~al.}(2024)\citenamefont {Pedersen}, \citenamefont {Lehtola}, \citenamefont {Fdez.~Galv{\'a}n},\ and\ \citenamefont {Lindh}}]{pedersen_etal_2024}%
  \BibitemOpen
  \bibfield  {author} {\bibinfo {author} {\bibfnamefont {T.~B.}\ \bibnamefont {Pedersen}}, \bibinfo {author} {\bibfnamefont {S.}~\bibnamefont {Lehtola}}, \bibinfo {author} {\bibfnamefont {I.}~\bibnamefont {Fdez.~Galv{\'a}n}},\ and\ \bibinfo {author} {\bibfnamefont {R.}~\bibnamefont {Lindh}},\ }\href {https://doi.org/10.1002/wcms.1692} {\bibfield  {journal} {\bibinfo  {journal} {WIREs Computational Molecular Science}\ }\textbf {\bibinfo {volume} {14}},\ \bibinfo {pages} {e1692} (\bibinfo {year} {2024})}\BibitemShut {NoStop}%
\bibitem [{\citenamefont {Motta}\ and\ \citenamefont {Zhang}(2018)}]{motta_zhang_2018}%
  \BibitemOpen
  \bibfield  {author} {\bibinfo {author} {\bibfnamefont {M.}~\bibnamefont {Motta}}\ and\ \bibinfo {author} {\bibfnamefont {S.}~\bibnamefont {Zhang}},\ }\href {https://doi.org/10.1002/wcms.1364} {\bibfield  {journal} {\bibinfo  {journal} {WIREs Computational Molecular Science}\ }\textbf {\bibinfo {volume} {8}},\ \bibinfo {pages} {e1364} (\bibinfo {year} {2018})}\BibitemShut {NoStop}%
\bibitem [{\citenamefont {Stratonovich}(1957)}]{stratonovich_1957}%
  \BibitemOpen
  \bibfield  {author} {\bibinfo {author} {\bibfnamefont {R.~L.}\ \bibnamefont {Stratonovich}},\ }\href@noop {} {\bibfield  {journal} {\bibinfo  {journal} {Soviet Physics Doklady}\ }\textbf {\bibinfo {volume} {2}},\ \bibinfo {pages} {416} (\bibinfo {year} {1957})}\BibitemShut {NoStop}%
\bibitem [{\citenamefont {Hubbard}(1959)}]{hubbard_1959}%
  \BibitemOpen
  \bibfield  {author} {\bibinfo {author} {\bibfnamefont {J.}~\bibnamefont {Hubbard}},\ }\href {https://doi.org/10.1103/PhysRevLett.3.77} {\bibfield  {journal} {\bibinfo  {journal} {Physical Review Letters}\ }\textbf {\bibinfo {volume} {3}},\ \bibinfo {pages} {77} (\bibinfo {year} {1959})}\BibitemShut {NoStop}%
\bibitem [{\citenamefont {Thouless}(1960)}]{thouless_1960}%
  \BibitemOpen
  \bibfield  {author} {\bibinfo {author} {\bibfnamefont {D.~J.}\ \bibnamefont {Thouless}},\ }\href {https://doi.org/10.1016/0029-5582(60)90048-1} {\bibfield  {journal} {\bibinfo  {journal} {Nuclear Physics}\ }\textbf {\bibinfo {volume} {21}},\ \bibinfo {pages} {225} (\bibinfo {year} {1960})}\BibitemShut {NoStop}%
\bibitem [{\citenamefont {Cholesky}(1924)}]{cholesky_1924}%
  \BibitemOpen
  \bibfield  {author} {\bibinfo {author} {\bibfnamefont {A.-L.}\ \bibnamefont {Cholesky}},\ }\href {https://doi.org/10.1007/BF03031308} {\bibfield  {journal} {\bibinfo  {journal} {Bulletin g\'eod\'esique}\ }\textbf {\bibinfo {volume} {2}},\ \bibinfo {pages} {67} (\bibinfo {year} {1924})}\BibitemShut {NoStop}%
\bibitem [{\citenamefont {Beebe}\ and\ \citenamefont {Linderberg}(1977)}]{beebe_linderberg_1977}%
  \BibitemOpen
  \bibfield  {author} {\bibinfo {author} {\bibfnamefont {N.~H.~F.}\ \bibnamefont {Beebe}}\ and\ \bibinfo {author} {\bibfnamefont {J.}~\bibnamefont {Linderberg}},\ }\href {https://doi.org/10.1002/qua.560120408} {\bibfield  {journal} {\bibinfo  {journal} {International Journal of Quantum Chemistry}\ }\textbf {\bibinfo {volume} {12}},\ \bibinfo {pages} {683} (\bibinfo {year} {1977})}\BibitemShut {NoStop}%
\bibitem [{\citenamefont {Koch}\ \emph {et~al.}(2003)\citenamefont {Koch}, \citenamefont {{S{\'a}nchez de Mer{\'a}s}},\ and\ \citenamefont {Pedersen}}]{koch_etal_2003}%
  \BibitemOpen
  \bibfield  {author} {\bibinfo {author} {\bibfnamefont {H.}~\bibnamefont {Koch}}, \bibinfo {author} {\bibfnamefont {A.}~\bibnamefont {{S{\'a}nchez de Mer{\'a}s}}},\ and\ \bibinfo {author} {\bibfnamefont {T.~B.}\ \bibnamefont {Pedersen}},\ }\href {https://doi.org/10.1063/1.1578621} {\bibfield  {journal} {\bibinfo  {journal} {The Journal of Chemical Physics}\ }\textbf {\bibinfo {volume} {118}},\ \bibinfo {pages} {9481} (\bibinfo {year} {2003})}\BibitemShut {NoStop}%
\bibitem [{\citenamefont {R{\o}eggen}\ and\ \citenamefont {Johansen}(2008)}]{roeggen_johansen_2008}%
  \BibitemOpen
  \bibfield  {author} {\bibinfo {author} {\bibfnamefont {I.}~\bibnamefont {R{\o}eggen}}\ and\ \bibinfo {author} {\bibfnamefont {T.}~\bibnamefont {Johansen}},\ }\href {https://doi.org/10.1063/1.2925269} {\bibfield  {journal} {\bibinfo  {journal} {The Journal of Chemical Physics}\ }\textbf {\bibinfo {volume} {128}},\ \bibinfo {pages} {194107} (\bibinfo {year} {2008})}\BibitemShut {NoStop}%
\bibitem [{\citenamefont {Aquilante}\ \emph {et~al.}(2010)\citenamefont {Aquilante}, \citenamefont {De~Vico}, \citenamefont {Ferr{\'e}}, \citenamefont {Ghigo}, \citenamefont {Malmqvist}, \citenamefont {Neogr{\'a}dy}, \citenamefont {Pedersen}, \citenamefont {Piton{\'a}k}, \citenamefont {Reiher}, \citenamefont {Roos}, \citenamefont {{Serrano-Andr{\'e}s}}, \citenamefont {Urban}, \citenamefont {Veryazov},\ and\ \citenamefont {Lindh}}]{aquilante_etal_2010}%
  \BibitemOpen
  \bibfield  {author} {\bibinfo {author} {\bibfnamefont {F.}~\bibnamefont {Aquilante}}, \bibinfo {author} {\bibfnamefont {L.}~\bibnamefont {De~Vico}}, \bibinfo {author} {\bibfnamefont {N.}~\bibnamefont {Ferr{\'e}}}, \bibinfo {author} {\bibfnamefont {G.}~\bibnamefont {Ghigo}}, \bibinfo {author} {\bibfnamefont {P.-{\aa}.}\ \bibnamefont {Malmqvist}}, \bibinfo {author} {\bibfnamefont {P.}~\bibnamefont {Neogr{\'a}dy}}, \bibinfo {author} {\bibfnamefont {T.~B.}\ \bibnamefont {Pedersen}}, \bibinfo {author} {\bibfnamefont {M.}~\bibnamefont {Piton{\'a}k}}, \bibinfo {author} {\bibfnamefont {M.}~\bibnamefont {Reiher}}, \bibinfo {author} {\bibfnamefont {B.~O.}\ \bibnamefont {Roos}}, \bibinfo {author} {\bibfnamefont {L.}~\bibnamefont {{Serrano-Andr{\'e}s}}}, \bibinfo {author} {\bibfnamefont {M.}~\bibnamefont {Urban}}, \bibinfo {author} {\bibfnamefont {V.}~\bibnamefont {Veryazov}},\ and\ \bibinfo {author} {\bibfnamefont {R.}~\bibnamefont {Lindh}},\ }\href {https://doi.org/10.1002/jcc.21318} {\bibfield  {journal} {\bibinfo
  {journal} {Journal of Computational Chemistry}\ }\textbf {\bibinfo {volume} {31}},\ \bibinfo {pages} {224} (\bibinfo {year} {2010})}\BibitemShut {NoStop}%
\bibitem [{\citenamefont {Hohenstein}\ \emph {et~al.}(2012{\natexlab{a}})\citenamefont {Hohenstein}, \citenamefont {Parrish},\ and\ \citenamefont {Mart{\'i}nez}}]{hohenstein_etal_2012}%
  \BibitemOpen
  \bibfield  {author} {\bibinfo {author} {\bibfnamefont {E.~G.}\ \bibnamefont {Hohenstein}}, \bibinfo {author} {\bibfnamefont {R.~M.}\ \bibnamefont {Parrish}},\ and\ \bibinfo {author} {\bibfnamefont {T.~J.}\ \bibnamefont {Mart{\'i}nez}},\ }\href {https://doi.org/10.1063/1.4732310} {\bibfield  {journal} {\bibinfo  {journal} {The Journal of Chemical Physics}\ }\textbf {\bibinfo {volume} {137}},\ \bibinfo {pages} {044103} (\bibinfo {year} {2012}{\natexlab{a}})}\BibitemShut {NoStop}%
\bibitem [{\citenamefont {Parrish}\ \emph {et~al.}(2012)\citenamefont {Parrish}, \citenamefont {Hohenstein}, \citenamefont {Mart{\'i}nez},\ and\ \citenamefont {Sherrill}}]{parrish_etal_2012}%
  \BibitemOpen
  \bibfield  {author} {\bibinfo {author} {\bibfnamefont {R.~M.}\ \bibnamefont {Parrish}}, \bibinfo {author} {\bibfnamefont {E.~G.}\ \bibnamefont {Hohenstein}}, \bibinfo {author} {\bibfnamefont {T.~J.}\ \bibnamefont {Mart{\'i}nez}},\ and\ \bibinfo {author} {\bibfnamefont {C.~D.}\ \bibnamefont {Sherrill}},\ }\href {https://doi.org/10.1063/1.4768233} {\bibfield  {journal} {\bibinfo  {journal} {The Journal of Chemical Physics}\ }\textbf {\bibinfo {volume} {137}},\ \bibinfo {pages} {224106} (\bibinfo {year} {2012})}\BibitemShut {NoStop}%
\bibitem [{\citenamefont {Hohenstein}\ \emph {et~al.}(2012{\natexlab{b}})\citenamefont {Hohenstein}, \citenamefont {Parrish}, \citenamefont {Sherrill},\ and\ \citenamefont {Mart{\'i}nez}}]{hohenstein_etal_2012a}%
  \BibitemOpen
  \bibfield  {author} {\bibinfo {author} {\bibfnamefont {E.~G.}\ \bibnamefont {Hohenstein}}, \bibinfo {author} {\bibfnamefont {R.~M.}\ \bibnamefont {Parrish}}, \bibinfo {author} {\bibfnamefont {C.~D.}\ \bibnamefont {Sherrill}},\ and\ \bibinfo {author} {\bibfnamefont {T.~J.}\ \bibnamefont {Mart{\'i}nez}},\ }\href {https://doi.org/10.1063/1.4768241} {\bibfield  {journal} {\bibinfo  {journal} {The Journal of Chemical Physics}\ }\textbf {\bibinfo {volume} {137}},\ \bibinfo {pages} {221101} (\bibinfo {year} {2012}{\natexlab{b}})}\BibitemShut {NoStop}%
\bibitem [{\citenamefont {Zhang}(2026)}]{Zhang:2026aa}%
  \BibitemOpen
  \bibfield  {author} {\bibinfo {author} {\bibfnamefont {Y.}~\bibnamefont {Zhang}},\ }\href {https://doi.org/10.1021/acs.jctc.5c01925} {\bibfield  {journal} {\bibinfo  {journal} {Journal of Chemical Theory and Computation}\ }\textbf {\bibinfo {volume} {22}},\ \bibinfo {pages} {1770} (\bibinfo {year} {2026})}\BibitemShut {NoStop}%
\bibitem [{\citenamefont {Trotter}(1959)}]{trotter_1959}%
  \BibitemOpen
  \bibfield  {author} {\bibinfo {author} {\bibfnamefont {H.~F.}\ \bibnamefont {Trotter}},\ }\href {https://doi.org/10.1090/S0002-9939-1959-0108732-6} {\bibfield  {journal} {\bibinfo  {journal} {Proceedings of the American Mathematical Society}\ }\textbf {\bibinfo {volume} {10}},\ \bibinfo {pages} {545} (\bibinfo {year} {1959})}\BibitemShut {NoStop}%
\bibitem [{\citenamefont {Suzuki}(1976)}]{suzuki_1976}%
  \BibitemOpen
  \bibfield  {author} {\bibinfo {author} {\bibfnamefont {M.}~\bibnamefont {Suzuki}},\ }\href {https://doi.org/10.1143/PTP.56.1454} {\bibfield  {journal} {\bibinfo  {journal} {Progress of Theoretical Physics}\ }\textbf {\bibinfo {volume} {56}},\ \bibinfo {pages} {1454} (\bibinfo {year} {1976})}\BibitemShut {NoStop}%
\bibitem [{\citenamefont {Zhang}\ \emph {et~al.}(1997)\citenamefont {Zhang}, \citenamefont {Carlson},\ and\ \citenamefont {Gubernatis}}]{zhang_etal_1997}%
  \BibitemOpen
  \bibfield  {author} {\bibinfo {author} {\bibfnamefont {S.}~\bibnamefont {Zhang}}, \bibinfo {author} {\bibfnamefont {J.}~\bibnamefont {Carlson}},\ and\ \bibinfo {author} {\bibfnamefont {J.~E.}\ \bibnamefont {Gubernatis}},\ }\href {https://doi.org/10.1103/PhysRevB.55.7464} {\bibfield  {journal} {\bibinfo  {journal} {Physical Review B}\ }\textbf {\bibinfo {volume} {55}},\ \bibinfo {pages} {7464} (\bibinfo {year} {1997})}\BibitemShut {NoStop}%
\bibitem [{\citenamefont {Loh}\ \emph {et~al.}(1990)\citenamefont {Loh}, \citenamefont {Gubernatis}, \citenamefont {Scalettar}, \citenamefont {White}, \citenamefont {Scalapino},\ and\ \citenamefont {Sugar}}]{loh_etal_1990}%
  \BibitemOpen
  \bibfield  {author} {\bibinfo {author} {\bibfnamefont {E.~Y.}\ \bibnamefont {Loh}}, \bibinfo {author} {\bibfnamefont {J.~E.}\ \bibnamefont {Gubernatis}}, \bibinfo {author} {\bibfnamefont {R.~T.}\ \bibnamefont {Scalettar}}, \bibinfo {author} {\bibfnamefont {S.~R.}\ \bibnamefont {White}}, \bibinfo {author} {\bibfnamefont {D.~J.}\ \bibnamefont {Scalapino}},\ and\ \bibinfo {author} {\bibfnamefont {R.~L.}\ \bibnamefont {Sugar}},\ }\href {https://doi.org/10.1103/PhysRevB.41.9301} {\bibfield  {journal} {\bibinfo  {journal} {Physical Review B}\ }\textbf {\bibinfo {volume} {41}},\ \bibinfo {pages} {9301} (\bibinfo {year} {1990})}\BibitemShut {NoStop}%
\bibitem [{\citenamefont {Ceperley}(1991)}]{ceperley_1991}%
  \BibitemOpen
  \bibfield  {author} {\bibinfo {author} {\bibfnamefont {D.~M.}\ \bibnamefont {Ceperley}},\ }\href {https://doi.org/10.1007/BF01030009} {\bibfield  {journal} {\bibinfo  {journal} {Journal of Statistical Physics}\ }\textbf {\bibinfo {volume} {63}},\ \bibinfo {pages} {1237} (\bibinfo {year} {1991})}\BibitemShut {NoStop}%
\bibitem [{\citenamefont {Zhang}\ and\ \citenamefont {Krakauer}(2003)}]{zhang_krakauer_2003}%
  \BibitemOpen
  \bibfield  {author} {\bibinfo {author} {\bibfnamefont {S.}~\bibnamefont {Zhang}}\ and\ \bibinfo {author} {\bibfnamefont {H.}~\bibnamefont {Krakauer}},\ }\href {https://doi.org/10.1103/PhysRevLett.90.136401} {\bibfield  {journal} {\bibinfo  {journal} {Physical Review Letters}\ }\textbf {\bibinfo {volume} {90}},\ \bibinfo {pages} {136401} (\bibinfo {year} {2003})}\BibitemShut {NoStop}%
\bibitem [{\citenamefont {Danilov}\ \emph {et~al.}(2025)\citenamefont {Danilov}, \citenamefont {Ganoe}, \citenamefont {Munyi},\ and\ \citenamefont {Shee}}]{danilov_etal_2025}%
  \BibitemOpen
  \bibfield  {author} {\bibinfo {author} {\bibfnamefont {D.}~\bibnamefont {Danilov}}, \bibinfo {author} {\bibfnamefont {B.}~\bibnamefont {Ganoe}}, \bibinfo {author} {\bibfnamefont {M.}~\bibnamefont {Munyi}},\ and\ \bibinfo {author} {\bibfnamefont {J.}~\bibnamefont {Shee}},\ }\href {https://doi.org/10.1021/acs.jctc.4c01251} {\bibfield  {journal} {\bibinfo  {journal} {Journal of Chemical Theory and Computation}\ }\textbf {\bibinfo {volume} {21}},\ \bibinfo {pages} {1136} (\bibinfo {year} {2025})}\BibitemShut {NoStop}%
\bibitem [{\citenamefont {Jiang}\ \emph {et~al.}(2025)\citenamefont {Jiang}, \citenamefont {O'Gorman}, \citenamefont {Mahajan},\ and\ \citenamefont {Lee}}]{jiang_etal_2025a}%
  \BibitemOpen
  \bibfield  {author} {\bibinfo {author} {\bibfnamefont {T.}~\bibnamefont {Jiang}}, \bibinfo {author} {\bibfnamefont {B.}~\bibnamefont {O'Gorman}}, \bibinfo {author} {\bibfnamefont {A.}~\bibnamefont {Mahajan}},\ and\ \bibinfo {author} {\bibfnamefont {J.}~\bibnamefont {Lee}},\ }\href {https://doi.org/10.1103/PhysRevResearch.7.013038} {\bibfield  {journal} {\bibinfo  {journal} {Physical Review Research}\ }\textbf {\bibinfo {volume} {7}},\ \bibinfo {pages} {013038} (\bibinfo {year} {2025})}\BibitemShut {NoStop}%
\bibitem [{\citenamefont {Mahajan}\ \emph {et~al.}(2025)\citenamefont {Mahajan}, \citenamefont {Thorpe}, \citenamefont {Kurian}, \citenamefont {Reichman}, \citenamefont {Matthews},\ and\ \citenamefont {Sharma}}]{mahajan_etal_2025}%
  \BibitemOpen
  \bibfield  {author} {\bibinfo {author} {\bibfnamefont {A.}~\bibnamefont {Mahajan}}, \bibinfo {author} {\bibfnamefont {J.~H.}\ \bibnamefont {Thorpe}}, \bibinfo {author} {\bibfnamefont {J.~S.}\ \bibnamefont {Kurian}}, \bibinfo {author} {\bibfnamefont {D.~R.}\ \bibnamefont {Reichman}}, \bibinfo {author} {\bibfnamefont {D.~A.}\ \bibnamefont {Matthews}},\ and\ \bibinfo {author} {\bibfnamefont {S.}~\bibnamefont {Sharma}},\ }\href {https://doi.org/10.1021/acs.jctc.4c01314} {\bibfield  {journal} {\bibinfo  {journal} {Journal of Chemical Theory and Computation}\ }\textbf {\bibinfo {volume} {21}},\ \bibinfo {pages} {1626} (\bibinfo {year} {2025})}\BibitemShut {NoStop}%
\bibitem [{\citenamefont {Xiao}\ \emph {et~al.}(2025{\natexlab{a}})\citenamefont {Xiao}, \citenamefont {Xiang}, \citenamefont {Lu}, \citenamefont {Chen},\ and\ \citenamefont {Zhang}}]{xiao_etal_2025}%
  \BibitemOpen
  \bibfield  {author} {\bibinfo {author} {\bibfnamefont {Z.-Y.}\ \bibnamefont {Xiao}}, \bibinfo {author} {\bibfnamefont {T.}~\bibnamefont {Xiang}}, \bibinfo {author} {\bibfnamefont {Z.}~\bibnamefont {Lu}}, \bibinfo {author} {\bibfnamefont {Y.}~\bibnamefont {Chen}},\ and\ \bibinfo {author} {\bibfnamefont {S.}~\bibnamefont {Zhang}},\ }\href {https://doi.org/10.1063/5.0296390} {\bibfield  {journal} {\bibinfo  {journal} {The Journal of Chemical Physics}\ }\textbf {\bibinfo {volume} {163}},\ \bibinfo {pages} {164109} (\bibinfo {year} {2025}{\natexlab{a}})}\BibitemShut {NoStop}%
\bibitem [{\citenamefont {Xiao}\ \emph {et~al.}(2025{\natexlab{b}})\citenamefont {Xiao}, \citenamefont {Kan}, \citenamefont {Ma}, \citenamefont {Zhao},\ and\ \citenamefont {Shang}}]{xiao_etal_2025a}%
  \BibitemOpen
  \bibfield  {author} {\bibinfo {author} {\bibfnamefont {Z.-Y.}\ \bibnamefont {Xiao}}, \bibinfo {author} {\bibfnamefont {B.}~\bibnamefont {Kan}}, \bibinfo {author} {\bibfnamefont {H.}~\bibnamefont {Ma}}, \bibinfo {author} {\bibfnamefont {B.}~\bibnamefont {Zhao}},\ and\ \bibinfo {author} {\bibfnamefont {H.}~\bibnamefont {Shang}},\ }\href {https://doi.org/10.1021/acs.jctc.5c01138} {\bibfield  {journal} {\bibinfo  {journal} {Journal of Chemical Theory and Computation}\ }\textbf {\bibinfo {volume} {21}},\ \bibinfo {pages} {9587} (\bibinfo {year} {2025}{\natexlab{b}})}\BibitemShut {NoStop}%
\bibitem [{\citenamefont {Yoshida}\ \emph {et~al.}(2025)\citenamefont {Yoshida}, \citenamefont {Erhart}, \citenamefont {Murokoshi}, \citenamefont {Nakagawa}, \citenamefont {Mori}, \citenamefont {Tagami},\ and\ \citenamefont {Mizukami}}]{yoshida_etal_2025}%
  \BibitemOpen
  \bibfield  {author} {\bibinfo {author} {\bibfnamefont {Y.}~\bibnamefont {Yoshida}}, \bibinfo {author} {\bibfnamefont {L.}~\bibnamefont {Erhart}}, \bibinfo {author} {\bibfnamefont {T.}~\bibnamefont {Murokoshi}}, \bibinfo {author} {\bibfnamefont {R.}~\bibnamefont {Nakagawa}}, \bibinfo {author} {\bibfnamefont {C.}~\bibnamefont {Mori}}, \bibinfo {author} {\bibfnamefont {H.}~\bibnamefont {Tagami}},\ and\ \bibinfo {author} {\bibfnamefont {W.}~\bibnamefont {Mizukami}},\ }\href {https://doi.org/10.1021/acs.jctc.5c00778} {\bibfield  {journal} {\bibinfo  {journal} {Journal of Chemical Theory and Computation}\ }\textbf {\bibinfo {volume} {21}},\ \bibinfo {pages} {9404} (\bibinfo {year} {2025})}\BibitemShut {NoStop}%
\bibitem [{\citenamefont {Williams}\ and\ \citenamefont {Zhang}(2026)}]{williams2026df}%
  \BibitemOpen
  \bibfield  {author} {\bibinfo {author} {\bibfnamefont {J.~Z.}\ \bibnamefont {Williams}}\ and\ \bibinfo {author} {\bibfnamefont {Y.}~\bibnamefont {Zhang}},\ }\href {https://arxiv.org/abs/2610.29851} {\bibinfo {title} {Differentiated determinant graphs enable large-scale multideterminant auxiliary-field quantum monte carlo}} (\bibinfo {year} {2026}),\ \Eprint {https://arxiv.org/abs/2610.29851} {arXiv:2610.29851 [physics.chem-ph]} \BibitemShut {NoStop}%
\bibitem [{\citenamefont {Wick}(1950)}]{wick_1950}%
  \BibitemOpen
  \bibfield  {author} {\bibinfo {author} {\bibfnamefont {G.~C.}\ \bibnamefont {Wick}},\ }\href {https://doi.org/10.1103/PhysRev.80.268} {\bibfield  {journal} {\bibinfo  {journal} {Physical Review}\ }\textbf {\bibinfo {volume} {80}},\ \bibinfo {pages} {268} (\bibinfo {year} {1950})}\BibitemShut {NoStop}%
\bibitem [{\citenamefont {Balian}\ and\ \citenamefont {Brezin}(1969)}]{balian_brezin_1969}%
  \BibitemOpen
  \bibfield  {author} {\bibinfo {author} {\bibfnamefont {R.}~\bibnamefont {Balian}}\ and\ \bibinfo {author} {\bibfnamefont {E.}~\bibnamefont {Brezin}},\ }\href {https://doi.org/10.1007/BF02710281} {\bibfield  {journal} {\bibinfo  {journal} {Il Nuovo Cimento B (1965-1970)}\ }\textbf {\bibinfo {volume} {64}},\ \bibinfo {pages} {37} (\bibinfo {year} {1969})}\BibitemShut {NoStop}%
\bibitem [{\citenamefont {Motta}\ \emph {et~al.}(2019)\citenamefont {Motta}, \citenamefont {Shee}, \citenamefont {Zhang},\ and\ \citenamefont {Chan}}]{motta_etal_2019a}%
  \BibitemOpen
  \bibfield  {author} {\bibinfo {author} {\bibfnamefont {M.}~\bibnamefont {Motta}}, \bibinfo {author} {\bibfnamefont {J.}~\bibnamefont {Shee}}, \bibinfo {author} {\bibfnamefont {S.}~\bibnamefont {Zhang}},\ and\ \bibinfo {author} {\bibfnamefont {G.~K.-L.}\ \bibnamefont {Chan}},\ }\href {https://doi.org/10.1021/acs.jctc.8b00996} {\bibfield  {journal} {\bibinfo  {journal} {Journal of Chemical Theory and Computation}\ }\textbf {\bibinfo {volume} {15}},\ \bibinfo {pages} {3510} (\bibinfo {year} {2019})}\BibitemShut {NoStop}%
\bibitem [{\citenamefont {{IEEE}}(2019)}]{ieee_2019}%
  \BibitemOpen
  \bibfield  {author} {\bibinfo {author} {\bibnamefont {{IEEE}}},\ }\href {https://doi.org/10.1109/IEEESTD.2019.8766229} {\bibinfo {title} {{{IEEE Standard}} for {{Floating-Point Arithmetic}}}} (\bibinfo {year} {2019})\BibitemShut {NoStop}%
\bibitem [{\citenamefont {{IEEE}}(1985)}]{ieee_1985}%
  \BibitemOpen
  \bibfield  {author} {\bibinfo {author} {\bibnamefont {{IEEE}}},\ }\href {https://doi.org/10.1109/IEEESTD.1985.82928} {\bibinfo {title} {{{IEEE Standard}} for {{Binary Floating-Point Arithmetic}}}} (\bibinfo {year} {1985})\BibitemShut {NoStop}%
\bibitem [{\citenamefont {Stosic}\ and\ \citenamefont {Micikevicius}(2021)}]{stosic_micikevicius_2021}%
  \BibitemOpen
  \bibfield  {author} {\bibinfo {author} {\bibfnamefont {D.}~\bibnamefont {Stosic}}\ and\ \bibinfo {author} {\bibfnamefont {P.}~\bibnamefont {Micikevicius}},\ }\href@noop {} {\bibinfo {title} {Accelerating {{AI Training}} with {{NVIDIA TF32 Tensor Cores}}}},\ \bibinfo {howpublished} {https://developer.nvidia.com/blog/accelerating-ai-training-with-tf32-tensor-cores/} (\bibinfo {year} {2021})\BibitemShut {NoStop}%
\bibitem [{nvi(2022)}]{nvidia_a100}%
  \BibitemOpen
  \href@noop {} {\bibinfo {title} {{{NVIDIA A100 GPUs Power}} the {{Modern Data Center}}}},\ \bibinfo {howpublished} {https://www.nvidia.com/en-us/data-center/a100/} (\bibinfo {year} {2022})\BibitemShut {NoStop}%
\bibitem [{\citenamefont {Steinbach}\ and\ \citenamefont {Bannwarth}(2025)}]{steinbach_bannwarth_2025}%
  \BibitemOpen
  \bibfield  {author} {\bibinfo {author} {\bibfnamefont {P.}~\bibnamefont {Steinbach}}\ and\ \bibinfo {author} {\bibfnamefont {C.}~\bibnamefont {Bannwarth}},\ }\href {https://doi.org/10.1021/acs.jctc.5c00262} {\bibfield  {journal} {\bibinfo  {journal} {Journal of Chemical Theory and Computation}\ }\textbf {\bibinfo {volume} {21}},\ \bibinfo {pages} {7335} (\bibinfo {year} {2025})}\BibitemShut {NoStop}%
\bibitem [{\citenamefont {Woo}\ and\ \citenamefont {Choi}(2026)}]{woo_choi_2026}%
  \BibitemOpen
  \bibfield  {author} {\bibinfo {author} {\bibfnamefont {J.}~\bibnamefont {Woo}}\ and\ \bibinfo {author} {\bibfnamefont {S.}~\bibnamefont {Choi}},\ }\href {https://doi.org/10.1021/acs.jctc.5c01800} {\bibfield  {journal} {\bibinfo  {journal} {Journal of Chemical Theory and Computation}\ }\textbf {\bibinfo {volume} {22}},\ \bibinfo {pages} {1687} (\bibinfo {year} {2026})}\BibitemShut {NoStop}%
\bibitem [{\citenamefont {{Olivares-Amaya}}\ \emph {et~al.}(2010)\citenamefont {{Olivares-Amaya}}, \citenamefont {Watson}, \citenamefont {Edgar}, \citenamefont {Vogt}, \citenamefont {Shao},\ and\ \citenamefont {{Aspuru-Guzik}}}]{olivares-amaya_etal_2010}%
  \BibitemOpen
  \bibfield  {author} {\bibinfo {author} {\bibfnamefont {R.}~\bibnamefont {{Olivares-Amaya}}}, \bibinfo {author} {\bibfnamefont {M.~A.}\ \bibnamefont {Watson}}, \bibinfo {author} {\bibfnamefont {R.~G.}\ \bibnamefont {Edgar}}, \bibinfo {author} {\bibfnamefont {L.}~\bibnamefont {Vogt}}, \bibinfo {author} {\bibfnamefont {Y.}~\bibnamefont {Shao}},\ and\ \bibinfo {author} {\bibfnamefont {A.}~\bibnamefont {{Aspuru-Guzik}}},\ }\href {https://doi.org/10.1021/ct900543q} {\bibfield  {journal} {\bibinfo  {journal} {Journal of Chemical Theory and Computation}\ }\textbf {\bibinfo {volume} {6}},\ \bibinfo {pages} {135} (\bibinfo {year} {2010})}\BibitemShut {NoStop}%
\bibitem [{\citenamefont {Eriksen}(2017)}]{eriksen_2017}%
  \BibitemOpen
  \bibfield  {author} {\bibinfo {author} {\bibfnamefont {J.~J.}\ \bibnamefont {Eriksen}},\ }\href {https://doi.org/10.1080/00268976.2016.1271155} {\bibfield  {journal} {\bibinfo  {journal} {Molecular Physics}\ }\textbf {\bibinfo {volume} {115}},\ \bibinfo {pages} {2086} (\bibinfo {year} {2017})}\BibitemShut {NoStop}%
\bibitem [{\citenamefont {Tian}\ \emph {et~al.}(2022)\citenamefont {Tian}, \citenamefont {Xie}, \citenamefont {Luo},\ and\ \citenamefont {Ma}}]{tian_etal_2022}%
  \BibitemOpen
  \bibfield  {author} {\bibinfo {author} {\bibfnamefont {Y.}~\bibnamefont {Tian}}, \bibinfo {author} {\bibfnamefont {Z.}~\bibnamefont {Xie}}, \bibinfo {author} {\bibfnamefont {Z.}~\bibnamefont {Luo}},\ and\ \bibinfo {author} {\bibfnamefont {H.}~\bibnamefont {Ma}},\ }\href {https://doi.org/10.1021/acs.jctc.2c00632} {\bibfield  {journal} {\bibinfo  {journal} {Journal of Chemical Theory and Computation}\ }\textbf {\bibinfo {volume} {18}},\ \bibinfo {pages} {7260} (\bibinfo {year} {2022})}\BibitemShut {NoStop}%
\bibitem [{\citenamefont {Huang}\ \emph {et~al.}(2024)\citenamefont {Huang}, \citenamefont {Guo}, \citenamefont {Pham},\ and\ \citenamefont {Lv}}]{huang_etal_2024}%
  \BibitemOpen
  \bibfield  {author} {\bibinfo {author} {\bibfnamefont {Y.}~\bibnamefont {Huang}}, \bibinfo {author} {\bibfnamefont {Z.}~\bibnamefont {Guo}}, \bibinfo {author} {\bibfnamefont {H.~Q.}\ \bibnamefont {Pham}},\ and\ \bibinfo {author} {\bibfnamefont {D.}~\bibnamefont {Lv}},\ }\href {https://doi.org/10.48550/arXiv.2406.08314} {\bibinfo {title} {{{GPU-accelerated Auxiliary-field}} quantum {{Monte Carlo}} with multi-{{Slater}} determinant trial states}} (\bibinfo {year} {2024}),\ \Eprint {https://arxiv.org/abs/2406.08314} {arXiv:2406.08314} \BibitemShut {NoStop}%
\bibitem [{\citenamefont {Thall}(2006)}]{thall_2006}%
  \BibitemOpen
  \bibfield  {author} {\bibinfo {author} {\bibfnamefont {A.}~\bibnamefont {Thall}},\ }in\ \href {https://doi.org/10.1145/1179622.1179682} {\emph {\bibinfo {booktitle} {{{ACM SIGGRAPH}} 2006 {{Research}} Posters}}},\ \bibinfo {series and number} {{{SIGGRAPH}} '06}\ (\bibinfo  {publisher} {Association for Computing Machinery},\ \bibinfo {address} {New York, NY, USA},\ \bibinfo {year} {2006})\ pp.\ \bibinfo {pages} {52--es}\BibitemShut {NoStop}%
\bibitem [{\citenamefont {{Valero-Lara}}\ \emph {et~al.}(2023)\citenamefont {{Valero-Lara}}, \citenamefont {Jorquera}, \citenamefont {Lui},\ and\ \citenamefont {Vetter}}]{valero-lara_etal_2023}%
  \BibitemOpen
  \bibfield  {author} {\bibinfo {author} {\bibfnamefont {P.}~\bibnamefont {{Valero-Lara}}}, \bibinfo {author} {\bibfnamefont {I.}~\bibnamefont {Jorquera}}, \bibinfo {author} {\bibfnamefont {F.}~\bibnamefont {Lui}},\ and\ \bibinfo {author} {\bibfnamefont {J.}~\bibnamefont {Vetter}},\ }in\ \href {https://doi.org/10.1145/3624062.3624084} {\emph {\bibinfo {booktitle} {Proceedings of the {{SC}} '23 {{Workshops}} of the {{International Conference}} on {{High Performance Computing}}, {{Network}}, {{Storage}}, and {{Analysis}}}}},\ \bibinfo {series and number} {{{SC-W}} '23}\ (\bibinfo  {publisher} {Association for Computing Machinery},\ \bibinfo {address} {New York, NY, USA},\ \bibinfo {year} {2023})\ pp.\ \bibinfo {pages} {179--186}\BibitemShut {NoStop}%
\bibitem [{\citenamefont {Fasi}\ \emph {et~al.}(2023)\citenamefont {Fasi}, \citenamefont {Higham}, \citenamefont {Lopez}, \citenamefont {Mary},\ and\ \citenamefont {Mikaitis}}]{fasi_etal_2023}%
  \BibitemOpen
  \bibfield  {author} {\bibinfo {author} {\bibfnamefont {M.}~\bibnamefont {Fasi}}, \bibinfo {author} {\bibfnamefont {N.~J.}\ \bibnamefont {Higham}}, \bibinfo {author} {\bibfnamefont {F.}~\bibnamefont {Lopez}}, \bibinfo {author} {\bibfnamefont {T.}~\bibnamefont {Mary}},\ and\ \bibinfo {author} {\bibfnamefont {M.}~\bibnamefont {Mikaitis}},\ }\href {https://doi.org/10.1137/21M1465032} {\bibfield  {journal} {\bibinfo  {journal} {SIAM Journal on Scientific Computing}\ }\textbf {\bibinfo {volume} {45}},\ \bibinfo {pages} {C1} (\bibinfo {year} {2023})}\BibitemShut {NoStop}%
\bibitem [{\citenamefont {Ozaki}\ \emph {et~al.}(2012)\citenamefont {Ozaki}, \citenamefont {Ogita}, \citenamefont {Oishi},\ and\ \citenamefont {Rump}}]{ozaki_etal_2012}%
  \BibitemOpen
  \bibfield  {author} {\bibinfo {author} {\bibfnamefont {K.}~\bibnamefont {Ozaki}}, \bibinfo {author} {\bibfnamefont {T.}~\bibnamefont {Ogita}}, \bibinfo {author} {\bibfnamefont {S.}~\bibnamefont {Oishi}},\ and\ \bibinfo {author} {\bibfnamefont {S.~M.}\ \bibnamefont {Rump}},\ }\href {https://doi.org/10.1007/s11075-011-9478-1} {\bibfield  {journal} {\bibinfo  {journal} {Numerical Algorithms}\ }\textbf {\bibinfo {volume} {59}},\ \bibinfo {pages} {95} (\bibinfo {year} {2012})}\BibitemShut {NoStop}%
\bibitem [{Note1()}]{Note1}%
  \BibitemOpen
  \bibinfo {note} {\protect \url {https://docs.nvidia.com/cuda/archive/12.9.0/cublas/index.html}}\BibitemShut {NoStop}%
\bibitem [{Note2()}]{Note2}%
  \BibitemOpen
  \bibinfo {note} {See \protect \url {https://developer.nvidia.com/blog/cublas-strided-batched-matrix-multiply/} and \protect \url {https://docs.nvidia.com/cuda/archive/12.9.0/cublas/index.html\#cublasgemmstridedbatchedex}; the moniker \protect \textsc {Ex} indicates that we can specify the precision.}\BibitemShut {Stop}%
\bibitem [{\citenamefont {Hehre}\ \emph {et~al.}(1969)\citenamefont {Hehre}, \citenamefont {Stewart},\ and\ \citenamefont {Pople}}]{hehre_etal_1969}%
  \BibitemOpen
  \bibfield  {author} {\bibinfo {author} {\bibfnamefont {W.~J.}\ \bibnamefont {Hehre}}, \bibinfo {author} {\bibfnamefont {R.~F.}\ \bibnamefont {Stewart}},\ and\ \bibinfo {author} {\bibfnamefont {J.~A.}\ \bibnamefont {Pople}},\ }\href {https://doi.org/10.1063/1.1672392} {\bibfield  {journal} {\bibinfo  {journal} {The Journal of Chemical Physics}\ }\textbf {\bibinfo {volume} {51}},\ \bibinfo {pages} {2657} (\bibinfo {year} {1969})}\BibitemShut {NoStop}%
\bibitem [{\citenamefont {Castellanos}(2026)}]{castellanos_2026}%
  \BibitemOpen
  \bibfield  {author} {\bibinfo {author} {\bibfnamefont {J.}~\bibnamefont {Castellanos}},\ }\href@noop {} {\bibinfo {title} {{{RayMol}}}} (\bibinfo {year} {2026}),\ \bibinfo {note} {\url{https://raymol.io}}\BibitemShut {NoStop}%
\bibitem [{\citenamefont {{Schr\"odinger, LLC}}(2025)}]{PyMOL}%
  \BibitemOpen
  \bibfield  {author} {\bibinfo {author} {\bibnamefont {{Schr\"odinger, LLC}}},\ }\href@noop {} {\bibinfo {title} {The {{PyMOL}} molecular graphics system, version 3.1.0}} (\bibinfo {year} {2025}),\ \bibinfo {note} {\url{https://pymol.org}}\BibitemShut {NoStop}%
\bibitem [{\citenamefont {Green}\ \emph {et~al.}(2021)\citenamefont {Green}, \citenamefont {Sly}, \citenamefont {Lapid}, \citenamefont {Ogas}, \citenamefont {Everson}, \citenamefont {Magee}, \citenamefont {Seamons},\ and\ \citenamefont {Ferrell}}]{green_etal_2021}%
  \BibitemOpen
  \bibfield  {author} {\bibinfo {author} {\bibfnamefont {J.~K.}\ \bibnamefont {Green}}, \bibinfo {author} {\bibfnamefont {N.~C.}\ \bibnamefont {Sly}}, \bibinfo {author} {\bibfnamefont {M.~F. T.~R.}\ \bibnamefont {Lapid}}, \bibinfo {author} {\bibfnamefont {J.~A.}\ \bibnamefont {Ogas}}, \bibinfo {author} {\bibfnamefont {K.~J.}\ \bibnamefont {Everson}}, \bibinfo {author} {\bibfnamefont {D.~J.}\ \bibnamefont {Magee}}, \bibinfo {author} {\bibfnamefont {C.~D.}\ \bibnamefont {Seamons}},\ and\ \bibinfo {author} {\bibfnamefont {P.~S.}\ \bibnamefont {Ferrell}},\ }in\ \href@noop {} {\emph {\bibinfo {booktitle} {{{CUG}}'21}}}\ (\bibinfo  {publisher} {Cray User Group},\ \bibinfo {address} {Los Alamos, New Mexico},\ \bibinfo {year} {2021})\BibitemShut {NoStop}%
\bibitem [{Note3()}]{Note3}%
  \BibitemOpen
  \bibinfo {note} {A stable version of \protect \textsc {OpenMS} is available at \protect \url {https://github.com/lanl/OpenMS}.}\BibitemShut {Stop}%
\bibitem [{\citenamefont {Sun}\ \emph {et~al.}(2018)\citenamefont {Sun}, \citenamefont {Berkelbach}, \citenamefont {Blunt}, \citenamefont {Booth}, \citenamefont {Guo}, \citenamefont {Li}, \citenamefont {Liu}, \citenamefont {McClain}, \citenamefont {Sayfutyarova}, \citenamefont {Sharma}, \citenamefont {Wouters},\ and\ \citenamefont {Chan}}]{sun_etal_2018}%
  \BibitemOpen
  \bibfield  {author} {\bibinfo {author} {\bibfnamefont {Q.}~\bibnamefont {Sun}}, \bibinfo {author} {\bibfnamefont {T.~C.}\ \bibnamefont {Berkelbach}}, \bibinfo {author} {\bibfnamefont {N.~S.}\ \bibnamefont {Blunt}}, \bibinfo {author} {\bibfnamefont {G.~H.}\ \bibnamefont {Booth}}, \bibinfo {author} {\bibfnamefont {S.}~\bibnamefont {Guo}}, \bibinfo {author} {\bibfnamefont {Z.}~\bibnamefont {Li}}, \bibinfo {author} {\bibfnamefont {J.}~\bibnamefont {Liu}}, \bibinfo {author} {\bibfnamefont {J.~D.}\ \bibnamefont {McClain}}, \bibinfo {author} {\bibfnamefont {E.~R.}\ \bibnamefont {Sayfutyarova}}, \bibinfo {author} {\bibfnamefont {S.}~\bibnamefont {Sharma}}, \bibinfo {author} {\bibfnamefont {S.}~\bibnamefont {Wouters}},\ and\ \bibinfo {author} {\bibfnamefont {G.~K.-L.}\ \bibnamefont {Chan}},\ }\href {https://doi.org/10.1002/wcms.1340} {\bibfield  {journal} {\bibinfo  {journal} {WIREs Computational Molecular Science}\ }\textbf {\bibinfo {volume} {8}},\ \bibinfo {pages} {e1340} (\bibinfo {year} {2018})}\BibitemShut
  {NoStop}%
\bibitem [{\citenamefont {Sun}\ \emph {et~al.}(2020)\citenamefont {Sun}, \citenamefont {Zhang}, \citenamefont {Banerjee}, \citenamefont {Bao}, \citenamefont {Barbry}, \citenamefont {Blunt}, \citenamefont {Bogdanov}, \citenamefont {Booth}, \citenamefont {Chen}, \citenamefont {Cui}, \citenamefont {Eriksen}, \citenamefont {Gao}, \citenamefont {Guo}, \citenamefont {Hermann}, \citenamefont {Hermes}, \citenamefont {Koh}, \citenamefont {Koval}, \citenamefont {Lehtola}, \citenamefont {Li}, \citenamefont {Liu}, \citenamefont {Mardirossian}, \citenamefont {McClain}, \citenamefont {Motta}, \citenamefont {Mussard}, \citenamefont {Pham}, \citenamefont {Pulkin}, \citenamefont {Purwanto}, \citenamefont {Robinson}, \citenamefont {Ronca}, \citenamefont {Sayfutyarova}, \citenamefont {Scheurer}, \citenamefont {Schurkus}, \citenamefont {Smith}, \citenamefont {Sun}, \citenamefont {Sun}, \citenamefont {Upadhyay}, \citenamefont {Wagner}, \citenamefont {Wang}, \citenamefont {White}, \citenamefont {Whitfield}, \citenamefont
  {Williamson}, \citenamefont {Wouters}, \citenamefont {Yang}, \citenamefont {Yu}, \citenamefont {Zhu}, \citenamefont {Berkelbach}, \citenamefont {Sharma}, \citenamefont {Sokolov},\ and\ \citenamefont {Chan}}]{sun_etal_2020}%
  \BibitemOpen
  \bibfield  {author} {\bibinfo {author} {\bibfnamefont {Q.}~\bibnamefont {Sun}}, \bibinfo {author} {\bibfnamefont {X.}~\bibnamefont {Zhang}}, \bibinfo {author} {\bibfnamefont {S.}~\bibnamefont {Banerjee}}, \bibinfo {author} {\bibfnamefont {P.}~\bibnamefont {Bao}}, \bibinfo {author} {\bibfnamefont {M.}~\bibnamefont {Barbry}}, \bibinfo {author} {\bibfnamefont {N.~S.}\ \bibnamefont {Blunt}}, \bibinfo {author} {\bibfnamefont {N.~A.}\ \bibnamefont {Bogdanov}}, \bibinfo {author} {\bibfnamefont {G.~H.}\ \bibnamefont {Booth}}, \bibinfo {author} {\bibfnamefont {J.}~\bibnamefont {Chen}}, \bibinfo {author} {\bibfnamefont {Z.-H.}\ \bibnamefont {Cui}}, \bibinfo {author} {\bibfnamefont {J.~J.}\ \bibnamefont {Eriksen}}, \bibinfo {author} {\bibfnamefont {Y.}~\bibnamefont {Gao}}, \bibinfo {author} {\bibfnamefont {S.}~\bibnamefont {Guo}}, \bibinfo {author} {\bibfnamefont {J.}~\bibnamefont {Hermann}}, \bibinfo {author} {\bibfnamefont {M.~R.}\ \bibnamefont {Hermes}}, \bibinfo {author} {\bibfnamefont {K.}~\bibnamefont {Koh}},
  \bibinfo {author} {\bibfnamefont {P.}~\bibnamefont {Koval}}, \bibinfo {author} {\bibfnamefont {S.}~\bibnamefont {Lehtola}}, \bibinfo {author} {\bibfnamefont {Z.}~\bibnamefont {Li}}, \bibinfo {author} {\bibfnamefont {J.}~\bibnamefont {Liu}}, \bibinfo {author} {\bibfnamefont {N.}~\bibnamefont {Mardirossian}}, \bibinfo {author} {\bibfnamefont {J.~D.}\ \bibnamefont {McClain}}, \bibinfo {author} {\bibfnamefont {M.}~\bibnamefont {Motta}}, \bibinfo {author} {\bibfnamefont {B.}~\bibnamefont {Mussard}}, \bibinfo {author} {\bibfnamefont {H.~Q.}\ \bibnamefont {Pham}}, \bibinfo {author} {\bibfnamefont {A.}~\bibnamefont {Pulkin}}, \bibinfo {author} {\bibfnamefont {W.}~\bibnamefont {Purwanto}}, \bibinfo {author} {\bibfnamefont {P.~J.}\ \bibnamefont {Robinson}}, \bibinfo {author} {\bibfnamefont {E.}~\bibnamefont {Ronca}}, \bibinfo {author} {\bibfnamefont {E.~R.}\ \bibnamefont {Sayfutyarova}}, \bibinfo {author} {\bibfnamefont {M.}~\bibnamefont {Scheurer}}, \bibinfo {author} {\bibfnamefont {H.~F.}\ \bibnamefont {Schurkus}},
  \bibinfo {author} {\bibfnamefont {J.~E.~T.}\ \bibnamefont {Smith}}, \bibinfo {author} {\bibfnamefont {C.}~\bibnamefont {Sun}}, \bibinfo {author} {\bibfnamefont {S.-N.}\ \bibnamefont {Sun}}, \bibinfo {author} {\bibfnamefont {S.}~\bibnamefont {Upadhyay}}, \bibinfo {author} {\bibfnamefont {L.~K.}\ \bibnamefont {Wagner}}, \bibinfo {author} {\bibfnamefont {X.}~\bibnamefont {Wang}}, \bibinfo {author} {\bibfnamefont {A.}~\bibnamefont {White}}, \bibinfo {author} {\bibfnamefont {J.~D.}\ \bibnamefont {Whitfield}}, \bibinfo {author} {\bibfnamefont {M.~J.}\ \bibnamefont {Williamson}}, \bibinfo {author} {\bibfnamefont {S.}~\bibnamefont {Wouters}}, \bibinfo {author} {\bibfnamefont {J.}~\bibnamefont {Yang}}, \bibinfo {author} {\bibfnamefont {J.~M.}\ \bibnamefont {Yu}}, \bibinfo {author} {\bibfnamefont {T.}~\bibnamefont {Zhu}}, \bibinfo {author} {\bibfnamefont {T.~C.}\ \bibnamefont {Berkelbach}}, \bibinfo {author} {\bibfnamefont {S.}~\bibnamefont {Sharma}}, \bibinfo {author} {\bibfnamefont {A.~Y.}\ \bibnamefont
  {Sokolov}},\ and\ \bibinfo {author} {\bibfnamefont {G.~K.-L.}\ \bibnamefont {Chan}},\ }\href {https://doi.org/10.1063/5.0006074} {\bibfield  {journal} {\bibinfo  {journal} {The Journal of Chemical Physics}\ }\textbf {\bibinfo {volume} {153}},\ \bibinfo {pages} {024109} (\bibinfo {year} {2020})}\BibitemShut {NoStop}%
\bibitem [{\citenamefont {Sun}\ \emph {et~al.}(2026)\citenamefont {Sun}, \citenamefont {Hermes}, \citenamefont {Wu}, \citenamefont {Zhai}, \citenamefont {Zhang}, \citenamefont {Ahmed}, \citenamefont {Aucar}, \citenamefont {Backhouse}, \citenamefont {Banerjee}, \citenamefont {Bao}, \citenamefont {Bogdanov}, \citenamefont {Bystrom}, \citenamefont {Chapoton}, \citenamefont {Chen}, \citenamefont {Chernyshov}, \citenamefont {Clifford}, \citenamefont {{Cohen-Janes}}, \citenamefont {Cui}, \citenamefont {Damour}, \citenamefont {Dattani}, \citenamefont {Dittmer}, \citenamefont {Ehlert}, \citenamefont {Eriksen}, \citenamefont {Evangelista}, \citenamefont {Ewing}, \citenamefont {Farahvash}, \citenamefont {Focke}, \citenamefont {Gao}, \citenamefont {Gasperich}, \citenamefont {Gillispie}, \citenamefont {Greiner}, \citenamefont {Hennefarth}, \citenamefont {Hermann}, \citenamefont {Hillenbrand}, \citenamefont {Huhtasalo}, \citenamefont {Ibrahim}, \citenamefont {Jangid}, \citenamefont {Javaremi}, \citenamefont {Jenkins},
  \citenamefont {Jin}, \citenamefont {King}, \citenamefont {Kooi}, \citenamefont {Kurian}, \citenamefont {Larsson}, \citenamefont {Lau}, \citenamefont {Lee}, \citenamefont {Lehtola}, \citenamefont {Li}, \citenamefont {Li}, \citenamefont {Li}, \citenamefont {Li}, \citenamefont {Li}, \citenamefont {Lykhin}, \citenamefont {Mahajan}, \citenamefont {Mauger}, \citenamefont {{del Mazo-Sevillano}}, \citenamefont {Moussa}, \citenamefont {Nakano}, \citenamefont {Neufeld}, \citenamefont {Peng}, \citenamefont {Pham}, \citenamefont {Pinski}, \citenamefont {Pokhilko}, \citenamefont {Pu}, \citenamefont {Qian}, \citenamefont {Quiton}, \citenamefont {Schulze}, \citenamefont {Scott}, \citenamefont {Seal}, \citenamefont {Serna}, \citenamefont {Smith}, \citenamefont {Smyser}, \citenamefont {Stahl}, \citenamefont {Sun}, \citenamefont {Sung}, \citenamefont {Trushin}, \citenamefont {Upadhyay}, \citenamefont {Vo}, \citenamefont {Vogels}, \citenamefont {Wang}, \citenamefont {Wang}, \citenamefont {Wang}, \citenamefont {Wang},
  \citenamefont {Wang}, \citenamefont {Williamson}, \citenamefont {Yang}, \citenamefont {Ye}, \citenamefont {Yeh}, \citenamefont {Yu}, \citenamefont {Yu}, \citenamefont {Yu}, \citenamefont {Zhang}, \citenamefont {Zhang}, \citenamefont {Zhang}, \citenamefont {Zhao}, \citenamefont {Zhou}, \citenamefont {Zhu}, \citenamefont {Zhu}, \citenamefont {Berkelbach}, \citenamefont {Gagliardi}, \citenamefont {Sharma}, \citenamefont {Sokolov},\ and\ \citenamefont {Chan}}]{sun_etal_2026}%
  \BibitemOpen
  \bibfield  {author} {\bibinfo {author} {\bibfnamefont {Q.}~\bibnamefont {Sun}}, \bibinfo {author} {\bibfnamefont {M.~R.}\ \bibnamefont {Hermes}}, \bibinfo {author} {\bibfnamefont {X.}~\bibnamefont {Wu}}, \bibinfo {author} {\bibfnamefont {H.}~\bibnamefont {Zhai}}, \bibinfo {author} {\bibfnamefont {X.}~\bibnamefont {Zhang}}, \bibinfo {author} {\bibfnamefont {A.~M.}\ \bibnamefont {Ahmed}}, \bibinfo {author} {\bibfnamefont {J.~J.}\ \bibnamefont {Aucar}}, \bibinfo {author} {\bibfnamefont {O.~J.}\ \bibnamefont {Backhouse}}, \bibinfo {author} {\bibfnamefont {S.}~\bibnamefont {Banerjee}}, \bibinfo {author} {\bibfnamefont {P.}~\bibnamefont {Bao}}, \bibinfo {author} {\bibfnamefont {N.~A.}\ \bibnamefont {Bogdanov}}, \bibinfo {author} {\bibfnamefont {K.}~\bibnamefont {Bystrom}}, \bibinfo {author} {\bibfnamefont {F.}~\bibnamefont {Chapoton}}, \bibinfo {author} {\bibfnamefont {N.-Y.}\ \bibnamefont {Chen}}, \bibinfo {author} {\bibfnamefont {I.~Y.}\ \bibnamefont {Chernyshov}}, \bibinfo {author} {\bibfnamefont {H.~S.}\
  \bibnamefont {Clifford}}, \bibinfo {author} {\bibfnamefont {S.}~\bibnamefont {{Cohen-Janes}}}, \bibinfo {author} {\bibfnamefont {Z.-H.}\ \bibnamefont {Cui}}, \bibinfo {author} {\bibfnamefont {Y.~D.}\ \bibnamefont {Damour}}, \bibinfo {author} {\bibfnamefont {N.}~\bibnamefont {Dattani}}, \bibinfo {author} {\bibfnamefont {L.~B.}\ \bibnamefont {Dittmer}}, \bibinfo {author} {\bibfnamefont {S.}~\bibnamefont {Ehlert}}, \bibinfo {author} {\bibfnamefont {J.~J.}\ \bibnamefont {Eriksen}}, \bibinfo {author} {\bibfnamefont {F.~A.}\ \bibnamefont {Evangelista}}, \bibinfo {author} {\bibfnamefont {S.~A.}\ \bibnamefont {Ewing}}, \bibinfo {author} {\bibfnamefont {A.}~\bibnamefont {Farahvash}}, \bibinfo {author} {\bibfnamefont {K.}~\bibnamefont {Focke}}, \bibinfo {author} {\bibfnamefont {Y.}~\bibnamefont {Gao}}, \bibinfo {author} {\bibfnamefont {K.~E.}\ \bibnamefont {Gasperich}}, \bibinfo {author} {\bibfnamefont {N.}~\bibnamefont {Gillispie}}, \bibinfo {author} {\bibfnamefont {J.}~\bibnamefont {Greiner}}, \bibinfo {author}
  {\bibfnamefont {M.~R.}\ \bibnamefont {Hennefarth}}, \bibinfo {author} {\bibfnamefont {J.}~\bibnamefont {Hermann}}, \bibinfo {author} {\bibfnamefont {C.}~\bibnamefont {Hillenbrand}}, \bibinfo {author} {\bibfnamefont {J.}~\bibnamefont {Huhtasalo}}, \bibinfo {author} {\bibfnamefont {B.}~\bibnamefont {Ibrahim}}, \bibinfo {author} {\bibfnamefont {B.}~\bibnamefont {Jangid}}, \bibinfo {author} {\bibfnamefont {A.~N.}\ \bibnamefont {Javaremi}}, \bibinfo {author} {\bibfnamefont {A.~J.}\ \bibnamefont {Jenkins}}, \bibinfo {author} {\bibfnamefont {Y.}~\bibnamefont {Jin}}, \bibinfo {author} {\bibfnamefont {D.~S.}\ \bibnamefont {King}}, \bibinfo {author} {\bibfnamefont {D.~P.}\ \bibnamefont {Kooi}}, \bibinfo {author} {\bibfnamefont {J.~S.}\ \bibnamefont {Kurian}}, \bibinfo {author} {\bibfnamefont {H.~R.}\ \bibnamefont {Larsson}}, \bibinfo {author} {\bibfnamefont {B.~T.~G.}\ \bibnamefont {Lau}}, \bibinfo {author} {\bibfnamefont {S.}~\bibnamefont {Lee}}, \bibinfo {author} {\bibfnamefont {S.}~\bibnamefont {Lehtola}},
  \bibinfo {author} {\bibfnamefont {C.}~\bibnamefont {Li}}, \bibinfo {author} {\bibfnamefont {H.}~\bibnamefont {Li}}, \bibinfo {author} {\bibfnamefont {J.}~\bibnamefont {Li}}, \bibinfo {author} {\bibfnamefont {R.}~\bibnamefont {Li}}, \bibinfo {author} {\bibfnamefont {S.}~\bibnamefont {Li}}, \bibinfo {author} {\bibfnamefont {A.~O.}\ \bibnamefont {Lykhin}}, \bibinfo {author} {\bibfnamefont {A.}~\bibnamefont {Mahajan}}, \bibinfo {author} {\bibfnamefont {N.}~\bibnamefont {Mauger}}, \bibinfo {author} {\bibfnamefont {P.}~\bibnamefont {{del Mazo-Sevillano}}}, \bibinfo {author} {\bibfnamefont {J.}~\bibnamefont {Moussa}}, \bibinfo {author} {\bibfnamefont {K.}~\bibnamefont {Nakano}}, \bibinfo {author} {\bibfnamefont {V.~A.}\ \bibnamefont {Neufeld}}, \bibinfo {author} {\bibfnamefont {L.}~\bibnamefont {Peng}}, \bibinfo {author} {\bibfnamefont {H.~Q.}\ \bibnamefont {Pham}}, \bibinfo {author} {\bibfnamefont {P.}~\bibnamefont {Pinski}}, \bibinfo {author} {\bibfnamefont {P.}~\bibnamefont {Pokhilko}}, \bibinfo {author}
  {\bibfnamefont {Z.}~\bibnamefont {Pu}}, \bibinfo {author} {\bibfnamefont {Y.}~\bibnamefont {Qian}}, \bibinfo {author} {\bibfnamefont {S.~J.}\ \bibnamefont {Quiton}}, \bibinfo {author} {\bibfnamefont {W.~T.}\ \bibnamefont {Schulze}}, \bibinfo {author} {\bibfnamefont {T.~R.}\ \bibnamefont {Scott}}, \bibinfo {author} {\bibfnamefont {A.}~\bibnamefont {Seal}}, \bibinfo {author} {\bibfnamefont {J.~D.}\ \bibnamefont {Serna}}, \bibinfo {author} {\bibfnamefont {J.~E.~T.}\ \bibnamefont {Smith}}, \bibinfo {author} {\bibfnamefont {K.~E.}\ \bibnamefont {Smyser}}, \bibinfo {author} {\bibfnamefont {T.}~\bibnamefont {Stahl}}, \bibinfo {author} {\bibfnamefont {C.}~\bibnamefont {Sun}}, \bibinfo {author} {\bibfnamefont {K.~J.}\ \bibnamefont {Sung}}, \bibinfo {author} {\bibfnamefont {E.}~\bibnamefont {Trushin}}, \bibinfo {author} {\bibfnamefont {S.}~\bibnamefont {Upadhyay}}, \bibinfo {author} {\bibfnamefont {E.~A.}\ \bibnamefont {Vo}}, \bibinfo {author} {\bibfnamefont {T.}~\bibnamefont {Vogels}}, \bibinfo {author}
  {\bibfnamefont {S.}~\bibnamefont {Wang}}, \bibinfo {author} {\bibfnamefont {T.}~\bibnamefont {Wang}}, \bibinfo {author} {\bibfnamefont {X.}~\bibnamefont {Wang}}, \bibinfo {author} {\bibfnamefont {X.}~\bibnamefont {Wang}}, \bibinfo {author} {\bibfnamefont {Y.}~\bibnamefont {Wang}}, \bibinfo {author} {\bibfnamefont {M.}~\bibnamefont {Williamson}}, \bibinfo {author} {\bibfnamefont {J.}~\bibnamefont {Yang}}, \bibinfo {author} {\bibfnamefont {H.-Z.}\ \bibnamefont {Ye}}, \bibinfo {author} {\bibfnamefont {C.-N.}\ \bibnamefont {Yeh}}, \bibinfo {author} {\bibfnamefont {H.}~\bibnamefont {Yu}}, \bibinfo {author} {\bibfnamefont {J.}~\bibnamefont {Yu}}, \bibinfo {author} {\bibfnamefont {V.~W.-z.}\ \bibnamefont {Yu}}, \bibinfo {author} {\bibfnamefont {C.}~\bibnamefont {Zhang}}, \bibinfo {author} {\bibfnamefont {D.}~\bibnamefont {Zhang}}, \bibinfo {author} {\bibfnamefont {Y.}~\bibnamefont {Zhang}}, \bibinfo {author} {\bibfnamefont {Z.}~\bibnamefont {Zhao}}, \bibinfo {author} {\bibfnamefont {Z.}~\bibnamefont {Zhou}},
  \bibinfo {author} {\bibfnamefont {A.~J.}\ \bibnamefont {Zhu}}, \bibinfo {author} {\bibfnamefont {T.}~\bibnamefont {Zhu}}, \bibinfo {author} {\bibfnamefont {T.~C.}\ \bibnamefont {Berkelbach}}, \bibinfo {author} {\bibfnamefont {L.}~\bibnamefont {Gagliardi}}, \bibinfo {author} {\bibfnamefont {S.}~\bibnamefont {Sharma}}, \bibinfo {author} {\bibfnamefont {A.}~\bibnamefont {Sokolov}},\ and\ \bibinfo {author} {\bibfnamefont {G.~K.-L.}\ \bibnamefont {Chan}},\ }\href@noop {} {\bibinfo {title} {The {{Python Simulations}} of {{Chemistry Framework}}: 10 years of an open-source quantum chemistry project}} (\bibinfo {year} {2026}),\ \Eprint {https://arxiv.org/abs/2603.14155} {arXiv:2603.14155 [physics.chem-ph]} \BibitemShut {NoStop}%
\bibitem [{\citenamefont {Okuta}\ \emph {et~al.}(2017)\citenamefont {Okuta}, \citenamefont {Unno}, \citenamefont {Nishino}, \citenamefont {Hido},\ and\ \citenamefont {Loomis}}]{okuta_etal_2017}%
  \BibitemOpen
  \bibfield  {author} {\bibinfo {author} {\bibfnamefont {R.}~\bibnamefont {Okuta}}, \bibinfo {author} {\bibfnamefont {Y.}~\bibnamefont {Unno}}, \bibinfo {author} {\bibfnamefont {D.}~\bibnamefont {Nishino}}, \bibinfo {author} {\bibfnamefont {S.}~\bibnamefont {Hido}},\ and\ \bibinfo {author} {\bibfnamefont {C.}~\bibnamefont {Loomis}},\ }in\ \href {http://learningsys.org/nips17/assets/papers/paper_16.pdf} {\emph {\bibinfo {booktitle} {Proceedings of Workshop on Machine Learning Systems (LearningSys) in The Thirty-first Annual Conference on Neural Information Processing Systems (NIPS)}}}\ (\bibinfo {year} {2017})\BibitemShut {NoStop}%
\bibitem [{\citenamefont {Malone}\ \emph {et~al.}(2023)\citenamefont {Malone}, \citenamefont {Mahajan}, \citenamefont {Spencer},\ and\ \citenamefont {Lee}}]{malone_etal_2023}%
  \BibitemOpen
  \bibfield  {author} {\bibinfo {author} {\bibfnamefont {F.~D.}\ \bibnamefont {Malone}}, \bibinfo {author} {\bibfnamefont {A.}~\bibnamefont {Mahajan}}, \bibinfo {author} {\bibfnamefont {J.~S.}\ \bibnamefont {Spencer}},\ and\ \bibinfo {author} {\bibfnamefont {J.}~\bibnamefont {Lee}},\ }\href {https://doi.org/10.1021/acs.jctc.2c00934} {\bibfield  {journal} {\bibinfo  {journal} {Journal of Chemical Theory and Computation}\ }\textbf {\bibinfo {volume} {19}},\ \bibinfo {pages} {109} (\bibinfo {year} {2023})}\BibitemShut {NoStop}%
\bibitem [{\citenamefont {Jiang}\ \emph {et~al.}(2024)\citenamefont {Jiang}, \citenamefont {Baumgarten}, \citenamefont {Loos}, \citenamefont {Mahajan}, \citenamefont {Scemama}, \citenamefont {Ung}, \citenamefont {Zhang}, \citenamefont {Malone},\ and\ \citenamefont {Lee}}]{jiang_etal_2024}%
  \BibitemOpen
  \bibfield  {author} {\bibinfo {author} {\bibfnamefont {T.}~\bibnamefont {Jiang}}, \bibinfo {author} {\bibfnamefont {M.~K.~A.}\ \bibnamefont {Baumgarten}}, \bibinfo {author} {\bibfnamefont {P.-F.}\ \bibnamefont {Loos}}, \bibinfo {author} {\bibfnamefont {A.}~\bibnamefont {Mahajan}}, \bibinfo {author} {\bibfnamefont {A.}~\bibnamefont {Scemama}}, \bibinfo {author} {\bibfnamefont {S.~F.}\ \bibnamefont {Ung}}, \bibinfo {author} {\bibfnamefont {J.}~\bibnamefont {Zhang}}, \bibinfo {author} {\bibfnamefont {F.~D.}\ \bibnamefont {Malone}},\ and\ \bibinfo {author} {\bibfnamefont {J.}~\bibnamefont {Lee}},\ }\href {https://doi.org/10.1063/5.0225596} {\bibfield  {journal} {\bibinfo  {journal} {The Journal of Chemical Physics}\ }\textbf {\bibinfo {volume} {161}},\ \bibinfo {pages} {162502} (\bibinfo {year} {2024})}\BibitemShut {NoStop}%
\end{thebibliography}%
